\documentclass[a4paper,11pt]{article}
\usepackage{jheppub}
\usepackage{amsthm}
\usepackage{tikz}
\usetikzlibrary{arrows.meta}
\usepackage{pgfplots}
\pgfplotsset{compat=1.18}
\usepgfplotslibrary{colormaps}
\usetikzlibrary{fit}
\usetikzlibrary{patterns}
\usepackage{braket}
\usepackage{cancel}
\usetikzlibrary{decorations.markings}
\usepackage{ytableau}
\ytableausetup{boxsize=0.5em}

\newcommand{\bZ}{\mathbb{Z}}

\usepackage{pict2e}
\makeatletter
\newcommand{\farsquare}[2]{#1\,{\mathpalette\far@square{#2}}}
\newcommand{\far@square}[2]{
  \mathop{\vcenter{\hbox{
    \sbox\z@{$\m@th#1\sum$}
    \setlength{\unitlength}{0.9\dimexpr\ht\z@+\dp\z@}
    \begin{picture}(1,1)
    \roundjoin
    \polyline(0,0)(0,1)(1,1)(1,0)(0,0)(0,0.5)
    \end{picture}
  }}}\limits_{#1#2}
}
\makeatother

\makeatletter

\newcommand{\cubeBC}[3]{\mathpalette\cubeBC@{{#1}{#2}{#3}}}

\newcommand{\cubeBC@}[2]{
\mathord{\vcenter{\hbox{
\setlength{\unitlength}{2.6ex} 
\begin{picture}(2.4,2.4)
\roundjoin

\polyline(0.35,0.35)(0.35,1.35)(1.35,1.35)(1.35,0.35)(0.35,0.35)
\polyline(0.80,0.80)(0.80,1.80)(1.80,1.80)(1.80,0.80)(0.80,0.80)
\polyline(0.35,0.35)(0.80,0.80)
\polyline(0.35,1.35)(0.80,1.80)
\polyline(1.35,1.35)(1.80,1.80)
\polyline(1.35,0.35)(1.80,0.80)

\put(0.18,0.85){\makebox(0,0)[r]{$\scriptstyle\firstofthree#2$}}
\put(0.85,0.18){\makebox(0,0)[t]{$\scriptstyle\secondofthree#2$}} 
\put(1.65,0.35){\makebox(0,0)[l]{$\scriptstyle\thirdofthree#2$}}  

\end{picture}
}}}
}

\newcommand{\firstofthree}[3]{#1}
\newcommand{\secondofthree}[3]{#2}
\newcommand{\thirdofthree}[3]{#3}

\makeatother

\usepackage{pgfplots}
\pgfplotsset{compat=1.18}
\usepackage{tikz}

\makeatletter
\def\@fpheader{}
\makeatother
\definecolor{thatorange}{RGB}{213, 94,0}
\definecolor{dark-blue2}{RGB}{0, 114,178}
\usepackage{hyperref}
\hypersetup{
	colorlinks=true,
	linkcolor=dark-blue2,
	citecolor=dark-blue2,
	urlcolor=dark-blue2,
	linktoc=section
}

\usepackage{tocloft}

\title{A Duality Web for Non-Supersymmetric Strings}

\author[1]{Zihni Kaan Baykara,}
\author[1,2]{Matilda Delgado,}
\author[3]{Emilian Dudas,}
\author[1]{Hector Parra De Freitas,}
\author[1]{Cumrun Vafa}

\affiliation[1]{Jefferson Physical Laboratory, Harvard University,\\
Cambridge, MA 02138, USA}
\affiliation[2]{Max-Planck-Institut f\"ur Physik (Werner-Heisenberg-Institut)\\
Boltzmannstr. 8, 85748 Garching, Germany}
\affiliation[3]{CPHT, CNRS,
Ecole polytechnique, Institut Polytechnique de Paris,\\
91120 Palaiseau, FRANCE}

\emailAdd{zbaykara@g.harvard.edu, matildadelgado@fas.harvard.edu, emilian.dudas@polytechnique.edu, hparradefreitas@fas.harvard.edu, vafa@g.harvard.edu}

\abstract{Motivated by the recently proposed geometric descriptions of 0A and 0B in M-theory and F-theory, we propose a web of duality among non-supersymmetric strings. In particular we argue that the distinct $\mathbb{Z}_2$ quotients of M-theory on $S^1\vee S^1$ lead to both 0A orientifolds as well as non-supersymmetric 10d heterotic vacua of the E-type, including the tachyon-free $SO(16)\times SO(16)$ strings.  Moreover we identify certain $\mathbb{Z}_2$ quotients of F-theory on $(S^1\vee S^1)\times S^1$ with 0B orientifolds (including a tachyon-free model) as well as others with dual to non-supersymmetric heterotic strings of the D-type.  Moreover using this picture we resolve some puzzles and provide further evidence for the Bergman-Gaberdiel duality between a particular 0B orientifold in 10 dimensions and the Narain compactification of 26-dimensional bosonic strings on a 16-dimensional torus, as well as the DMS conjecture of a 0A orientifold duality in 10d with a bosonic string orientifold of a Narain compactification to 10d.
}
\begin{document}
\maketitle
\flushbottom

\section{Introduction}

Superstring dualities have played an important role in deepening our understanding of non-perturbative phenomena in quantum gravitational theories that enjoy supersymmetry. The same cannot be said, unfortunately, about quantum gravity theories without supersymmetry.  It is natural to ask whether non-trivial dualities also apply to this class to shed light on their strong coupling dynamics.  There have already been some conjectured strong-weak coupling dualities for non-supersymmetric strings \cite{Bergman:1997rf,Blum:1997cs,Blum:1997gw,Antoniadis:1998ki,Kachru:1998yy,Blumenhagen:1999ad,Bergman:1999km,Fabinger:2000jd,Dudas:2001wd,Dudas:2004nd,Hellerman:2004zm,Angelantonj:2007ts,Faraggi:2007tj,Acharya:2022shu,bcd,Fraiman:2025yrx}. However, unlike their supersymmetric counterparts, generically there are unresolved puzzles for these to work. Moreover, there are very few tests that can be performed on these dualities, as the lack of supersymmetry can lead to strong quantum corrections that cannot be controlled as we go from weak to strong coupling. Furthermore, they typically have tachyons and it is natural to ask what the implications of tachyon condensations are in these theories.\footnote{For reviews on string theories without supersymmetry, see e.g. \cite{nonsusystringsrevs1,Mourad:2017rrl,nonsusystringsrevs2,nonsusystringsrevs3,nonsusystringsrevs4}.
For other issues related to classical and quantum stability of non-supersymmetric vacua, see e.g. \cite{Angelantonj:2006ut,Basile:2018irz,Antonelli:2019nar,Basile:2021vxh,Mourad:2021roa,Raucci:2022bjw,Mourad:2024mpg}.}  

Already in 10 dimensions we have a number of non-supersymmetric strings, some of which have tachyons whilst others do not. Perhaps the simplest ones are those obtained by non-supersymmetric orbifolds of IIA and IIB strings by $(-1)^F$ projecting out the fermions, leading to 0A and 0B superstrings.  Recently M-theory and F-theory lifts of these theories were proposed on $S^1\vee S^1$ and $(S^1\vee S^1)\times S^1$ respectively \cite{Baykara:2026gem}.  It is natural to ask whether other non-supersymmetric strings fit in this framework.
The main aim of this work is to show that they also form a non-trivial web of dualities.  In particular we focus on $\bZ_2$ quotients of M-theory and F-theory descriptions of 0A and 0B and argue that they include all the non-supersymmetric 10d heterotic strings as well as 0A and 0B orientifolds\footnote{The general notion of an orientifold was introduced in \cite{augusto}.}.

\begin{figure}
    \centering
    \includegraphics[width=1\linewidth]{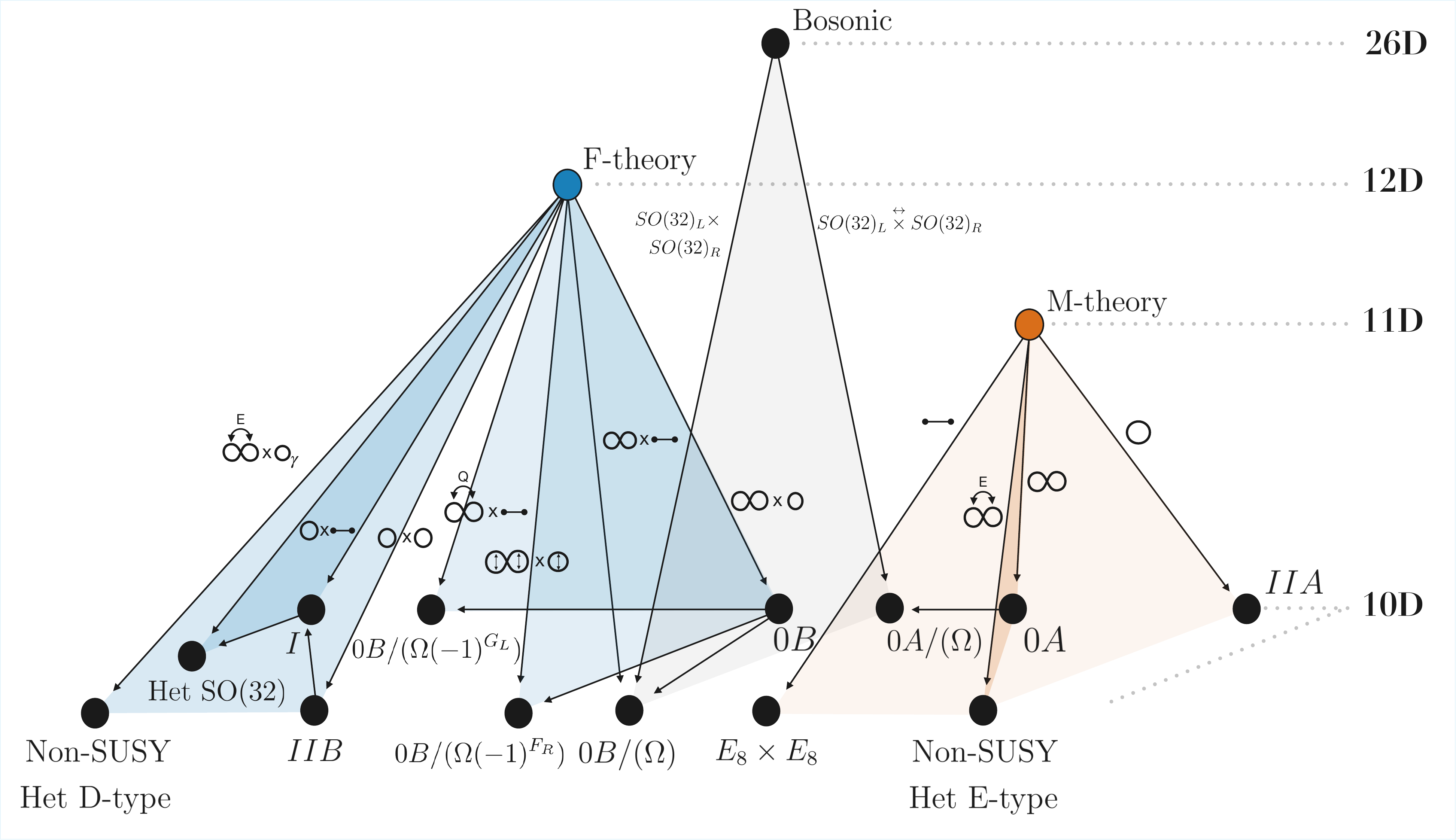}
    \caption{Schematic overview of the network of non-supersymmetric dualities discussed in this work. The dualities connect all ten-dimensional heterotic string theories and perturbative orientifold constructions of Type 0A and Type 0B strings, and to M-theory, F-theory and bosonic string theory.  The labels on the arrows specify the relevant compactifications of M and F-theory.}
    \label{fig:dualitymap}
\end{figure}

One of the non-supersymmetric 0B orientifolds with $SO(32)\times SO(32)$ gauge symmetry has been conjectured by Bergman and Gaberdiel \cite{Bergman:1997rf} to be dual to the 26-dimensional bosonic string compactified on a Narain lattice corresponding to $SO(32)_L\otimes SO(32)_R$ gauge symmetry.  While they found some evidence for this duality, they also point out issues with it including a mismatch of the field content.  We propose resolutions to those mismatches and provide additional evidence for this duality.  We thus provide what we consider to be strong evidence for the remarkable duality conjecture of BG connecting a non-supersymmetric string orientifold in 10d to compactifications of 26-dimensional bosonic strings!  Similarly, one of the 0A orientifolds with $SO(32)$ gauge symmetry has been conjectured in \cite{Dudas:2001wd} to be dual to the orientifold of the same bosonic string.  Here there are also similar puzzles and we show that the same method that we employ to resolve the BG puzzles can also resolve the mismatch for this proposed duality.\footnote{
There exist alternative constructions in the literature that relate critical, and non-critical string theories via space-time varying tachyon condensation \cite{Hellerman:2006ff,Hellerman:2006nx,Hellerman:2007fc,Kaidi:2020jla}. In contrast, we aim to preserve Lorentz invariance on both sides of the correspondence, allowing for a consistent matching of spectra as described above.  In this sense, our goal is to match the theories themselves, rather than focusing on specific background configurations or near-horizon geometries of extended objects \cite{Kaidi:2024cbx,Altavista:2026edv,Anastasi:2026cus,Angius:2022mgh}. Our results are not, \emph{a priori}, in contradiction with these other approaches to tachyon condensation. Nevertheless, it would be interesting to better understand how these different perspectives may fit together in future work.}

Let us briefly clarify the methods we employ when discussing these non-supersymmetric dualities. In the absence of supersymmetry, there are no BPS-protected quantities to guide the analysis, and masses are not protected against quantum corrections, except for massless $p$-form gauge fields (with $p>1$) which are protected by Lorentz invariance to exist, unless they are massed up by a Higgs mechanism (or decouple from the theory in another way as in \cite{Baykara:2026gem}) which requires the existence of a transition point as we go from weak to strong coupling. Chiral fermions should of course match on both sides (as long as the gauge field is not Higgsed). When apparent mismatches arise for massless gauge fields, we regard it as necessary to provide a reasonable physical explanation such transition points.   That there always are candidates for such transition points is what we view as evidence for these non-supersymmetric dualities.

Finally, we highlight a key takeaway from this work: tachyons and their condensation play a central role in non-supersymmetric dualities. Indeed, typically when we want to go to strong coupling we need not only to increase string coupling but also to take a particular trajectory for the tachyon condensate. Because of this, tachyons are crucial in understanding dualities between Type 0A/0B orientifolds and bosonic string theory. Moreover, in M-theory and F-theory constructions, the tachyon admits a geometric realization, and tachyon condensation can be interpreted in geometric terms.

The organization of this paper is as follows:
In \autoref{sec:S1S1-quotients} we discuss $\bZ_2$ quotients of M-theory on $S^1\vee S^1$ in 10d and connect them, depending on whether the $\bZ_2$ exchanges the circles or reflects one of them, to non-supersymmetric heterotic strings of the E-type or 0A orientifolds respectively. In \autoref{sec:F-theory} we discuss 0B $\bZ_2$ quotients of F-theory on $(S^1\vee S^1)\times S^1$.  We argue that when it reflects the $S^1$, without acting on $S^1\vee S^1$ it gives the various 0B orientifolds.  When it acts in addition to exchanging the two circles of $S^1\vee S^1$ it leads to the tachyon-free non-supersymmetric string of \cite{Alvarez-Gaume:1986ghj,Dixon:1986iz} and when it acts in addition to reflecting both circles of $S^1\vee S^1$ it leads to an orientifold of 0B with $\Omega (-1)^{F_R}$.  When it does not act on the $S^1$ but switches the two circles of $S^1\vee S^1$ we argue that these are dual to non-supersymmetric $SO$ heterotic strings.
When it acts by switching the two circles preserving the orientation of the circles but reflects the stand-alone circle, it corresponds to the $U(32)$ tachyon-free Sagnotti model \cite{Sagnotti:1995ga} corresponding to $0B/
\Omega Q$ where $Q$ is the quantum symmetry of 0B. In \autoref{sec:dualitiesbosonic} we discuss and provide further evidence for the Bergman-Gaberdiel duality between a 0B orientifold with 32 $D9^+$ and 32 $\overline{D9}^+$ branes and a particular bosonic string
compactified on the $SO(32)_L\otimes SO(32)_R$ Narain lattice.  We also provide further evidence for the proposed duality between a 0A orientifold with the orientifold of the bosonic string \cite{Dudas:2001wd} by suggesting how the field mismatch is resolved.  We present our concluding thoughts in \autoref{sec:conclude}.

While finishing this paper, another paper appeared \cite{Altavista:2026evd} which independently studies some of the theories considered here.  Our conclusions have some overlaps and some differences.

\section{M-theory $\bZ_2$ quotients on $S^1\vee S^1$}\label{sec:S1S1-quotients}
In this section we consider $\mathbb{Z}_2$ symmetries of M-theory on $S^1\vee S^1$ and the effect of quotienting the theory by them. We will consider four such symmetries, three are given by reflections of the two circles, and one is given by the exchange of them. We will argue that quotienting by two of them (reflection of one of the two circles) gives rise to an M-theoretic description of the perturbative orientifold of the 0A string. The remaining two quotients will instead correspond to E-type heterotic strings (meaning heterotic strings with  E-gauge symmetry or SO with a massless spinor), leading to all such non-supersymmetric cases that have been constructed.

We consider first the two symmetries $P^+$ and $P^-$ which act by reflection of either of the two circles in $S^1\vee S^1$. These are defined by
\begin{align}
	P^-: ~~~
	(\theta^+,\theta^-) &\mapsto (\theta^+,-\theta^-)\,,\nonumber\\\nonumber\\
	P^+: ~~~
	(\theta^+,\theta^-) &\mapsto (-\theta^+,\theta^-)\,.\nonumber
\end{align}
The action of $P^\pm$ on the RR and NSNS $p$-form fields in Type 0A is 
\begin{equation}\label{eq:Paction}
	\begin{split}
    \begin{array}{c|c}
        P^+ &  P^-\\ \hline
        A^\pm \to \mp A^\pm & A^\pm \to \pm A^\pm\\
        C^\pm \to \mp C^\pm & C^\pm \to \pm C^\pm\\
        B\to -B & B\to -B\\
    \end{array}
	\end{split}
\end{equation}
The action on the $A^\pm,C^\pm$ is clear from the 0A construction in terms of M-theory on $S^1 \vee S^1$ \cite{Baykara:2026gem}. The action on $B$ is less obvious, because $B$ lives on both circles. For this action to make sense it is important that $B$ has the strong smoothness property (SSP) and admits both the continuous resolution property (CRP) and its counterpart with opposite orientation on the second circle (CRP$'$).  The involution leads to exchanging the equivalence class of B-field represented by (CRP)$\leftrightarrow$(CRP$'$).  The fact that $B$ has to pick up a $-$ sign under this exchange is forced by the symmetries of 0A: the only symmetry that trades the twisted RR fields with the untwisted ones is the exchange of left/right movers which flips the sign of $B$.  We leave it as an open question to explain why this sign on the $B$ field is also demanded from the M-theoretical $S^1\vee S^1$ perspective.

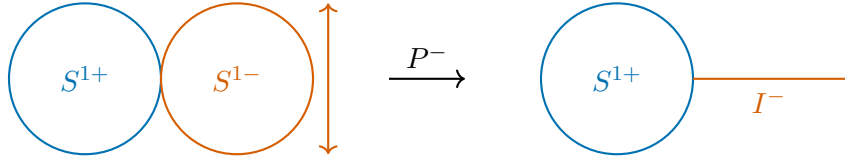
\begin{figure}[h!]
    \centering
    \vspace{0.3cm}
    \begin{tikzpicture}
		\draw[thick,dark-blue2](0,0) circle (1) node{$S^{1+}$};
		\draw[thick,thatorange](2,0) circle (1) node{$S^{1-}$};
		\draw[thick](4,0)--(4.5,0)node[above]{$P^-$};
		\draw[thick, ->](4.5,0)--(5,0);
        \draw[thatorange,<->, thick] (3.2,-1) -- (3.2,1);
		\begin{scope}[shift={(7,0)}]
			\draw[thick,dark-blue2](0,0) circle (1) node{$S^{1+}$};
			\draw[thick,thatorange](1,0)--(2,0) node[below]{$I^-$};
            \draw[thick,thatorange](2,0)--(3,0);
		\end{scope}
	\end{tikzpicture} 
    \caption{Quotienting $S^{1+}\vee S^{1-}$ by a reflection $P^{-}$ on the second circle $S^{1-}$ results in a circle joined with an interval $S^{1+}\vee I^{-}$.}
    \label{fig:Pminusquotient}
      \vspace{0.1cm}
\end{figure}

Quotienting $S^1\vee S^1$ by $P^-$ or $P^+$ we obtain the geometries $S^{1+}\vee I^-$ and $I^+\vee S^{1-}$, where the wedge point on the interval sits at a boundary. The first is depicted in \autoref{fig:Pminusquotient}.

\noindent We denote the corresponding quotients by
\begin{equation}
\begin{aligned}
P^- &:\quad S^{1+} \,\vee\, \bigl(S^{1-} \updownarrow\bigr), \\
P^+ &:\quad \bigl(S^{1+} \updownarrow\bigr) \,\vee\, S^{1-}.
\end{aligned}
\end{equation}
We propose that M-theory on these geometries correspond to the two (physically equivalent) orientifolds of the Type 0A string, which we discuss in \autoref{sec:0Aorient}.

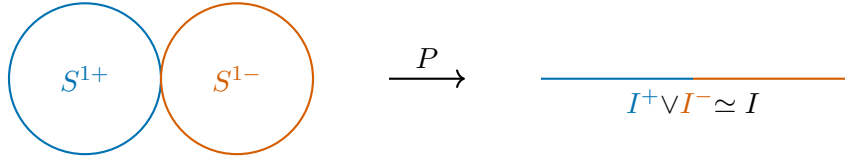
\begin{figure}[h!]
  \vspace{0.3cm}
    \centering
    \begin{tikzpicture}
		\draw[thick,dark-blue2](0,0) circle (1) node{$S^{1+}$};
		\draw[thick,thatorange](2,0) circle (1) node{$S^{1-}$};
		\draw[thick](4,0)--(4.5,0)node[above]{$P$};
		\draw[thick, ->](4.5,0)--(5,0);
		\begin{scope}[shift={(7,0)}]
			\draw[thick,dark-blue2](-1,0)--(1,0) node[below]{$\textcolor{dark-blue2}{I^+}\textcolor{black}{\vee} \textcolor{thatorange}{I^-} \textcolor{black}{\simeq I}$};
            \draw[thick,thatorange](1,0)--(3,0);
		\end{scope}
	\end{tikzpicture}    
    \caption{The quotient of $S^1 \vee S^1$ by $P \equiv P^+P^-$ leads to an interval $I^+\vee I^- \simeq I$.}
    \label{fig:intervalquotient}
      \vspace{0.1cm}
\end{figure}

Next we consider the combination $P \equiv P^+P^-$. This symmetry projects out all the RR fields, but preserves the NSNS B-field. Geometrically, this quotient leads to an interval $I^+\vee I^- \simeq I$, as depicted in \autoref{fig:intervalquotient}. 

\noindent We denote this quotient by
\begin{align}
    (S^{1+} \vee S^{1-})\updownarrow.
\end{align}
Note that whenever there are multiple $\updownarrow$ symbols, the quotient is taken by the simultaneous action of the reflections.  We will propose that this quotient results in the usual Ho\v{r}ava-Witten compactification of M-theory describing the supersymmetric $E_8\times E_8$ heterotic string in \autoref{sec:e8e8}. 

Finally we consider the exchange action of the two circles.  Note that there are two possibilities for this.  One exchange keeps the orientations of the two circles (corresponding to a half shift on the covering circle) which is the quantum symmetry $Q$.  Quotienting by this gives back the IIA theory, as was already noted in \cite{Baykara:2026gem}.  Here we are interested in the other exchange $E$ which reverses orientations of the circles
\begin{align}
    E: \ \qquad  S^{1+}\overset{\overset{E}\leftrightarrow}\vee S^{1-}
\end{align}
This can also be viewed in the connected resolution as leading to two boundaries that are stuck together:  
\begin{equation}
\tikzset{every picture/.style={line width=0.75pt}}
\begin{tikzpicture}[x=0.75pt,y=0.75pt,yscale=1,xscale=1]
\tikzset{
  mydraw/.style={draw,line width=0.75pt,line cap=round,line join=round},
  arr/.style={->,line width=0.75pt,line cap=round}
}
\pgfdeclarelayer{bg}
\pgfdeclarelayer{fg}
\pgfsetlayers{bg,main,fg}
\def\a{50}   
\def\b{50}   
\def\cx{0}
\def\cy{0}
\def\cxx{200}
\draw[mydraw]
  plot[domain=0:360,samples=300]
    ({\cx + 35 * cos(\x)*sqrt(1^2*cos(2*\x)+sqrt(1.01^4-1^4 *sin(2*\x)^2))},
     {-\cy + \a * sin(\x)*sqrt(1^2*cos(2*\x)+sqrt(1.01^4-1^4 *sin(2*\x)^2))});

\draw[mydraw] (175,3) circle (25);

\draw[white,line width=3pt] (199.2,5.2) -- (199.2,0.8);

\draw[thick,dark-blue2] (190.5,5.2) -- (210,5.2);
\draw[thick,dark-blue2] (190.5,0.8) -- (210,0.8);
  
\draw[<->, thick]
  (-12,20) to[out=30,in=150] (12,20);

\draw[mydraw] (25,120) circle (25);
\draw[mydraw] (-25,120) circle (25);

\draw[mydraw] (175,120) circle (25);

\draw[black,fill = black](0,6.5) circle (1.5); 
\draw[black,fill = black](0,-6.5) circle (1.5);
\draw[thick,dotted](0,6.5)--(0,-6.5);

\draw[thick,dark-blue2](190,120)--(210,120); 
\draw[thick](80,120)--(100,120)node[above]{$E$};
\draw[thick, ->](100,120)--(120,120);
\draw[thick](80,0)--(100,0)node[below]{$E$};
\draw[thick, ->](100,0)--(120,0);

\draw[thick](0,80)--(0,60)node[left]{CRP};
\draw[thick, ->](0,60)--(0,40);
\draw[thick](175,80)--(175,60)node[right]{DBRP};
\draw[thick, ->](175,60)--(175,40);
\end{tikzpicture}
\end{equation}
The dashed lines signify that the two points cannot be physically pulled apart, and DBRP stands for disconnected boundary resolution property. Note that unlike the quantum symmetry, which acts as a rotation on the connected resolution, $E$ is a reflection and has fixed points. We will be identifying the resulting quotient with the E-type heterotic strings in \autoref{sec:e-type-het}.

\subsection{0A orientifold: $S^1\vee (S^1\updownarrow$)}\label{sec:0Aorient}

We first review the construction of the Type 0A orientifold \cite{bianchi-sagnotti} (an extended discussion of this orientifold can be found in \cite{Dudas:2001wd}) and explain which of its properties can be described in terms of the geometry
\begin{align}
    S^1\vee (S^1\updownarrow)\; = S^1\vee I\,,
\end{align}
and what more we can learn from it. 

Unlike the Type IIA string, worldsheet parity is a symmetry of the Type 0A string without having to reflect one of the spatial coordinates. This is easily seen from its torus partition function 
\begin{equation}\label{Z0A}
	\mathcal{T}_{0A} \sim V_8 \bar V_8 + O_8 \bar O_8 + S_8 \bar C_8 +  C_8 \bar S_8\,.
\end{equation}
This theory can be constructed as an orbifold of the Type IIA string by $(-1)^F$, with $F$ the spacetime fermion number. From this perspective, the term $S_8 \bar C_8$ ($C_8 \bar S_8$) corresponds to the RR fields in the untwisted (twisted) sector. We denote the untwisted and twisted RR fields respectively as $A, C$ and $A',C'$. Worldsheet parity trades left-movers with right-movers, and from its action on \eqref{Z0A} it is easy to see that it exchanges the twisted and untwisted RR sectors. It is helpful to define the linear combinations:
\begin{equation}
    A^\pm = A \pm A'\,, ~~~~~ C^\pm = C \pm C'  \,.
\end{equation}
Worldsheet parity acts on these fields by preserving $A^+,C^+$ and projecting out $A^-,C^-$. The action on the NSNS sectors is the standard orientifold one, keeping the symmetric left-right combinations, namely the metric $g$, the dilaton $\phi$ and the tachyon $T_{0A}$. 

The orientifold of the Type 0A string is not T-dual to any ten-dimensional orientifold of the Type 0B string,\footnote{The T-dual theory will involve a compactification on an interval. We focus on theories with 10d Lorentz invariance.} thus we denote it by $\Omega_{0A}^-$, where the superscript refers to its non-trivial action on $A^-,C^-$. It is clear that there exists an alternative but physically equivalent orientifold action $\Omega_{0A}^+$ given by 
\begin{equation}
	\Omega_{0A}^+ = \Omega_{0A}^- (-1)^{F_R^s}\,,
\end{equation}
where $F_R^s$ is the right-moving spacetime fermion number, hence $(-1)^{F_R^s}$ acts as $-1$ on the RR sectors, cf.  eq. \eqref{Z0A}. Equivalently, $\Omega_{0A}^\pm$ is the conjugate of $\Omega_{0A}^\mp$ by the quantum symmetry $Q$. We then propose that these orientifold actions lift to M-theory as
\begin{equation}
    \Omega_{0A}^\pm \to P^\pm\,.
\end{equation}
Hence, the non-oriented Type 0A strings are described by M-theory on the geometries $S^1\vee I$ and $ I\vee S^1$. Indeed, one can check that the action of $\Omega_{0A}^\pm$ on the closed string sector of Type 0A agrees with the actions we discussed for $P^\pm$ in \eqref{eq:Paction}.

Let us now attempt to extend this correspondence to the open string sector. The standard Klein bottle computation gives rise to a negative dilaton tadpole, but no RR tadpole. The dilaton tadpole does not lead to an inconsistency, but we can choose to cancel it by adding a total of 32 D9-branes to the background. As usual, these branes come in two species $D9^+$ and $D9^-$, but in the Type 0A theory they have no RR charge, they descend from the non-BPS D9-branes of the Type IIA theory. We can add $n$ $D9^+$-branes and $(32-n)$ $D9^-$-branes, making the gauge group
\begin{equation}
    G = SO(n)\times SO(32-n)\,.
\end{equation}
The $D9^+$ and $D9^-$ branes differ in the sign of their coupling to the closed string tachyon. Before the orientifold action, their worldvolumes carry $U(n)$ gauge groups and tachyonic scalars in the adjoint representation. The standard Möbius strip computation shows that for $D9^-$, the orientifold $\Omega_{0A}^-$ projects the $U(32-n)$ adjoint scalars to the rank 2 symmetric representation $\ydiagram{2}$ of $SO(32-n)$, whereas for $D9^+$ we obtain the antisymmetric (adjoint) representation $\vcenter{\hbox{\ydiagram{1,1}}}$ of $SO(n)$.  Both are tachyonic fields.

This orientifold also has a non-zero tadpole for the closed string tachyon, which is absent for $n = 0$ \cite{Dudas:2001wd}\footnote{The sum of Klein bottle, annulus and Möbius strip contributions cancel in this case (there is a minor typo in \cite{Dudas:2001wd}).} and it was proposed there that this special case could be S-dual to an orientifold of the bosonic string compactified on the $SO(32)$ torus down to 10 dimensions, to which we will return in \autoref{sec:DMS}. Finally, there are generically massless (non-chiral) Majorana fermions in the bifundamental $(\Box,\Box)$ of $G$ (which are absent when $n=0$).

To explain the appearance of the $SO(n)$ gauge groups from the $S^1\vee I$ perspective, the presence of the interval $I$ suggests placing M9-branes at its boundaries. Recall, however, that the Type 0A $D9^\pm$-branes descend from the Type IIA non-BPS $D9$-branes, and these have been proposed in \cite{Houart:2000vm} to descend from non-BPS M10-branes in M-theory\footnote{It is tempting to view M10 as D10, i.e., the continuation of the even D-branes of IIA D0,...,D8 to include D10, except that M10 is not supersymmetric.} on $S^1$. From the perspective of M-theory on $S^1\vee S^1$ it is then more natural that the $D9^\pm$-branes correspond to M-theory M10-branes wrapped on $S^{1+}$ and $S^{1-}$ respectively.

Consider then a stack of $n_+$ M10-branes wrapped around $S^{1+}$ and $n_-$ M10-branes wrapped around $S^{1-}$. These become $D9^\pm$-branes once wrapped on a circle, whose worldvolume gauge groups are respectively $U(n_+)$ and $U(n_-)$. The orientifold action acts on the Chan-Paton degrees of freedom of each stack of $D9^\pm$ branes, projecting the gauge groups to $SO(n_+)$ and $SO(n_-)$. We also have tachyons in the adjoint representations of $U(n_+)$ and $U(n_-)$, which get projected respectively to the anti-symmetric and symmetric representations of the $SO(n_+)$ and $SO(n_-)$. We may then ask whether these projections can be explained from the point of view of M-theory on $S^1\vee S^1$.  A natural explanation is that the $P^-$ involution acts on both stacks of branes as it usually acts on D-branes \eqref{eq:Paction}.  For the ones wrapping $S^{1-}$, it is clear that $P^-$ acts non-trivially.  For the ones wrapping $S^{1+}$ we also need to act on the orientation of the circle. This is related to the fact that as we mentioned the $B$-field which lives on both circles has to flip sign $B\rightarrow -B$ and the orientation of branes thus has to be changed on both circles, leading to $SO(n_+)\times SO(n_-)$.  However, for the components of the gauge fields in the 11th direction, since one circle gets flipped, there is an extra sign compared to the action on the rest of the gauge field.  Note that the $U(n)$ adjoint breaks to $SO(n)$ representations $\ydiagram{2}\oplus \vcenter{\hbox{\ydiagram{1,1}}}$ (where we are including the singlet in the symmetric part).  
So for the $SO(n_+)$ all components $(A^+_\mu, A^+_{11})$ transform the same way as $(\vcenter{\hbox{\ydiagram{1,1}}},\vcenter{\hbox{\ydiagram{1,1}}})$, but for the $SO(n_-)$, due to the $P^-$ action on the $S^{1-}$, the components $(A^-_\mu,A^-_{11}) $ transform as $(\vcenter{\hbox{\ydiagram{1,1}}}, \ydiagram{2})$, and we identify the 11th component holonomies with the pair of scalar tachyons of the 0A orientifold, respectively in the correct order.  To complete the multiplet we need to assume that bifundamental fields of $U(n_+)\times U(n_-)$, which arise from the intersection of $M10^\pm$ branes, should descend to bifundamental fermions ($\ydiagram{1},\ydiagram{1})$ of $SO(n_+)\times SO(n_-)$ under the $P^-$ involution.

Finally, we can ask what the fate of tachyon condensation is in this theory. We have two inequivalent routes depending on how we put the branes. We may consider either the limit $R^+ \to 0$ or $R^- \to 0$. The second is straightforward, leaving out one circle, and we conjecture that this limit (at least when $n_+=0$) is just the supersymmetric Type IIA string (or decays to it if $n_+\not =0$).  The other limit is perhaps more interesting, and we will argue it is strong-weak dual to an orientifold of the bosonic string when $(n_+=0,n_-=32)$ in \autoref{sec:DMS} as has been conjectured in \cite{Dudas:2001wd}.

\subsection{Supersymmetric $E_8\times E_8$ string: $(S^1\vee S^1)\updownarrow$}\label{sec:e8e8}

We now consider the combined action of $\Omega_{0A}^+$ and $\Omega_{0A}^-$, which corresponds to the geometric action
\begin{align}
    P: \theta^\pm \mapsto -\theta^\pm.
\end{align}
We denote the quotient by
\begin{align}
    (S^1\vee S^1)\updownarrow,
\end{align}
where $\updownarrow$ is to be understood as a reflection on both circles, done simultaneously in a single $\bZ_2$ action. 
On the continuous resolution circle, it also acts as a reflection
\begin{align}
    P: \theta \mapsto -\theta.
\end{align}
One may wonder if there is something special about the wedge point of the two circles after quotienting by this symmetry. To clarify this problem we take a different route that will also be instructive for the case of the trading of the two circles by reordering the operations leading to this background\footnote{For this to be valid, we implicitly assume certain field resolution properties for the wedge sum which may differ from the one leading to 0A.  We will assume the field resolutions are chosen so that we can reorder the two operations.}.  Namely, we can think of the wedge as identifying two points on a circle and then acting with the involutions.  Reversing this order would mean we first do the involution on the circle and then identify the two points.  After the first step, we have an interval.  For the second step, the $P^\pm$ involution identifies a point on the interval with itself and thus does nothing.  In other words we would be left with the usual Ho\v{r}ava-Witten realization of $E_8\times E_8$. 

This suggests that modding by $P=P^+P^-$ in $S^1\vee S^1$ simply gives M-theory on the interval, which is equivalent to the $E_8 \times E_8$ heterotic string.

\subsection{E-type heterotic strings: $S^1\overset{\overset{E}\leftrightarrow}{\vee}S^1$} \label{sec:e-type-het}
\begin{table}[h!]
    \centering
    \begin{tabular}{|c|c|c|}\hline
         Gauge group & Tachyons & Fermions \\\hline
         \multicolumn{3}{|c|}{E-type} \\\hline
         $E_8$ & $(\mathbf{1})$ &
         $\begin{matrix}
             (\mathbf{248})_+\\
             (\mathbf{248})_-
         \end{matrix}$ \\\hline
         $SO(16)\times SO(16)$ & --- &
         $\begin{matrix}
             (\mathbf{128,1})_+\\
             (\mathbf{1,128})_+\\
             (\mathbf{16,16})_-
         \end{matrix}$ \\\hline
         $E_7\times SU(2)\times E_7\times SU(2)$ & $(\mathbf{1,2,1,2})$ &
         $\begin{matrix}
             (\mathbf{56,2,1,1})_+\\
             (\mathbf{1,1,56,2})_+\\
             (\mathbf{56,1,1,2})_-\\
             (\mathbf{1,2,56,1})_-
         \end{matrix}$ \\\hline
         $E_8\times SO(16)$ & $(\mathbf{1,16})$ &
         $\begin{matrix}
             (\mathbf{1,128})_+\\
             (\mathbf{1,128'})_-
         \end{matrix}$ \\\hline
         \multicolumn{3}{|c|}{D-type} \\\hline
         $SO(32)$ & $(\mathbf{32})$ & --- \\\hline
         $SO(24)\times SO(8)$ & $(\mathbf{1,8})$ &
         $\begin{matrix}
             (\mathbf{24,8_v})_+\\
             (\mathbf{24,8_c})_-
         \end{matrix}$ \\\hline
         $SU(16)\times U(1)$ & $(\mathbf{1},\pm 1)$ &
         $\begin{matrix}
             (\mathbf{120},+2)_+\\
             (\mathbf{\overline{120}},-2)_+\\
             (\mathbf{120},-2)_-\\
             (\mathbf{\overline{120}},+2)_-
         \end{matrix}$ \\\hline
    \end{tabular}
    \caption{Massless and tachyonic field content of the seven non-supersymmetric heterotic strings in 10D, organized into E-type and D-type classes. The E-type strings correspond to those with a gauge group that can be obtained from $E_8\times E_8$, whereas the D-type strings correspond to gauge groups obtained from $SO(32)$. Note that we also classify $SO(16)\times SO(16)$ as E-type because it has fermions in the spinor representation of $SO(16)$.}
    \label{tab:nonsusy_heterotic_10d}
\end{table}

There is another reflection action on the $S^1\vee S^1$ which exchanges the two circles,
\begin{align}
    E: \theta^+ \leftrightarrow \theta^-.
\end{align}
On the covering connected resolution circle $\hat S^1$, it acts as
\begin{align}
    E: \theta \mapsto \pi-\theta,
\end{align}
where $\theta =\pm\pi/2$ are the joining points of the circle. This action fixes the joining point of $S^1\vee S^1$. The resulting geometry therefore is an $S^1$ with a single quantum boundary fixed at a point.

\begin{figure}
     \centering
     \begin{tikzpicture}[x=0.75pt,y=0.75pt,yscale=1,xscale=1]
\tikzset{
  mydraw/.style={draw,line width=0.75pt,line cap=round,line join=round},
  arr/.style={->,line width=0.75pt,line cap=round}
}
\pgfdeclarelayer{bg}
\pgfdeclarelayer{fg}
\pgfsetlayers{bg,main,fg}

\draw[mydraw] (0,120) circle (40);

\draw[black,fill = black](0,160) circle (1.5); 
\draw[black,fill = black](0,80) circle (1.5);
\draw[thick,dotted](0,160)--(0,120)node[right]{$\vee$};
\draw[thick,dotted](0,120)--(0,80);

\draw[mydraw] (175,95) -- (175,145);

\draw[thick,dark-blue2] (165,145)--(185,145);
\draw[thick,dark-blue2] (165,95)--(185,95);
\draw[black,fill = black](175,145) circle (1.5); 
\draw[black,fill = black](175,95) circle (1.5);

\draw[mydraw] (-25,3) circle (25);
\draw[mydraw] (25,3) circle (25);

\draw[black,fill = black](0,3) circle (1.5);

\draw[<->, thick]
  (-10,165) to[out=30,in=150]
  node[pos=0.5, above] {$E$}
  (10,165);

\draw[mydraw] (175,3) circle (25);

\draw[white,line width=3pt] (199.2,5.2) -- (199.2,0.8);

\draw[thick,dark-blue2] (190.5,5.2) -- (210,5.2);
\draw[thick,dark-blue2] (190.5,0.8) -- (210,0.8);

\draw[thick](80,120)--(100,120) node[above]{$E$};
\draw[thick, ->](100,120)--(120,120);

\draw[thick](80,0)--(100,0) node[below]{$E$};
\draw[thick, ->](100,0)--(120,0);

\draw[thick](0,75)--(0,60) node[left]{$\vee$};
\draw[thick, ->](0,60)--(0,35);

\draw[thick](175,80)--(175,60) node[right]{$\vee$};
\draw[thick, ->](175,60)--(175,35);

\end{tikzpicture}
     \caption{The two routes to obtaining E-type heterotic strings. The bottom route corresponds to first identifying $\frac \pi 2 \sim - \frac \pi 2$ on the covering $\hat S^1$ and obtaining $S^1\vee S^1$, then exchanging the two circles by $E$ and obtaining $S^1$ with the quantum boundary. The top route corresponds to quotienting by $E$ first to get the Ho\v{r}ava-Witten interval, then identifying the boundaries.}
     \label{fig:type0AvsHW}
 \end{figure}
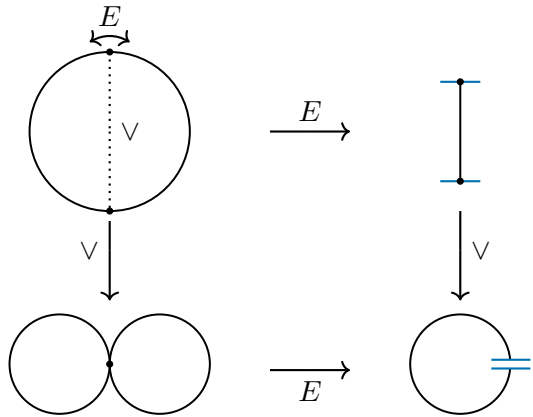

Similarly to the $E_8\times E_8$ case considered previously, it is useful to reverse the order of the operations and identify the joining points after the $\bZ_2^E$ involution of M-theory on the covering  $\hat S^1$, namely
\begin{align}
    \frac{S^1\vee S^1}{\bZ_2^E} = \Big[\frac{\hat S^1}{\bZ_2^E}\Big]\Big/(\pi/2 \sim -\pi/2 ).
\end{align}
The identification points are the two boundaries, corresponding to bringing the $E_8$ boundaries together in the geometry, see \autoref{fig:type0AvsHW}.

The non-supersymmetric heterotic strings are listed in \autoref{tab:nonsusy_heterotic_10d}. E-type strings correspond to the non-supersymmetric descendants of the $E_8\times E_8$ string, and D-type strings correspond to those of the $SO(32)$ string. Although $SO(16)\times SO(16)$ can be obtained from either, we classify it as an E-type string since it has fermions in the spinor representation of the $SO(16)$ factors, whereas fermions of the D-type strings have brane-like representations. In this section, we geometrize all the E-type heterotic strings with the above construction and will later geometrize the D-type heterotic strings in \autoref{sec:D-type}.

\subsubsection{${E_{8,2}}$} There is a worldsheet motivation for identifying the boundaries of the interval as well. The non-supersymmetric $E_{8,2}$ heterotic string is obtained by orbifolding the supersymmetric $E_8\times E_8$ string by the exchange of $E_8\leftrightarrow E_8$ accompanied by $(-1)^F$. This perturbative orbifold action is very suggestive that the M-theory lift corresponds to identifying the two boundaries together without acting on other points of the circle.  This has already been proposed in \cite{Faraggi:2007tj}, which becomes more meaningful in the setup we have.

The geometry corresponds to a circle with a frozen boundary singularity where the $E_{8,2}$ degrees of freedom live. The gauge fields in this case have the connected boundary resolution property (CBRP) in analogy with the CRP fields of 0A. Therefore we only get one copy of the gauge field living on the boundary. We thus argue that the M-theory lift of the non-supersymmetric $E_{8,2}$ heterotic string corresponds to a circle with a quantum boundary on which $E_{8,2}$ gauge degrees of freedom live.

\subsubsection{${SO(16)\times SO(16)}, {(E_7\times SU(2))^2}$}

It is conceivable that the quantum boundary on the circle follows the disconnected boundary resolution property (DBRP) in which case there are two copies of the gauge field. In the Ho\v{r}ava-Witten picture, this resolution corresponds to bringing the two boundaries close together, but not identifying them. This way, we have two copies of the gauge fields $A_\mu^1,A_\mu^2$, similar to how the DRP property produces two copies of gauge fields $A^\pm,C^\pm$ for Type 0A. 

There can also be fields that correspond to excitations in between the two boundaries which are now inseparable. In particular, these excitations can now be massless given that the boundaries are physically not separated.

In assuming DRP for the boundaries, microscopic consistency conditions may require breaking $E_8$ to a subgroup. Indeed, we find that all the E-type heterotic strings with DBRP have two gauge groups that each come from $E_8$.

Indeed, the spectrum of the non-supersymmetric tachyon-free $SO(16)\times SO(16)$ theory matches the prescription: there are two copies of the gauge field and the chiral fermions are stuck to the quantum boundary. The $(\mathbf{1,128})_+, (\mathbf{128,1})_+$ fermions come from breaking of $E_8$ on each boundary, and the $(\mathbf{16,16})_-$ fermion corresponds to a stretched mode in between.

In particular, the $(\mathbf{16,16})_-$ fermion is evidence that the two branes are stuck together and cannot be smoothly separated. This is because $(\mathbf{16,16})_-$ cancels the anomaly due to the fermions of opposite chirality that only live on one boundary. If the boundaries could be separated smoothly, then the $(\mathbf{16,16})_-$ fermion would mass up and the chiral anomaly from $(\mathbf{128,1})_+$ and $(\mathbf{1,128})_+$ fermions would not be canceled. Indeed the chiral field $(\mathbf{16,16})_-$ cannot be massed up, in line with the fact that boundaries cannot be macroscopically separated.  That bringing the boundary points close together could be a way to obtain the $SO(16)\times SO(16)$ was also proposed in \cite{Faraggi:2007tj}; here we have a natural explanation of what bringing them together is, and why they cannot be separated.

Similarly, the non-supersymmetric $E_7\times SU(2)\times E_7\times SU(2)$ heterotic string has two copies of gauge fields coming from each $E_8$, consistent with DBRP. There are fermions that live on each brane $(\mathbf{56,2,1,1})$ and also cross-brane $(\mathbf{56,1,1,2})$ made possible by the two boundaries being stuck together and provides evidence that the boundaries cannot be separated smoothly, as predicted by our picture.

Perhaps not unrelatedly, $SO(16)\times SO(16)$ and $E_7 \times SU(2) \times E_7 \times SU(2)$ are the only non-supersymmetric $E$-type heterotic string theories appearing in \autoref{tab:nonsusy_heterotic_10d} for which local anomaly cancellation requires the Green--Schwarz mechanism. Indeed, their 12-dimensional anomaly polynomials \cite{Alvarez-Gaume:1984zlq} factorize as:
\begin{equation}\label{eq:factorize}
    P^{SO(16)^2}_{12} \sim X^{SO(16)^2}_4 X^{SO(16)^2}_8\,,\quad P^{(E_7\times SU(2))^2}_{12} \sim X^{(E_7\times SU(2))^2}_4 X^{(E_7\times SU(2))^2}_8\,
\end{equation}
where 
\begin{align}
    X^{SO(16)^2}_4 &= (c_{2,1}+c_{2,2}+p_1)\,,\\
     X^{SO(16)^2}_8 &= (c_{2,1}^2+c_{2,2} c_{2,1}+c_{2,2}^2-4 \left(c_{4,1}+c_{4,2}\right))\,,
\end{align}
where $c_{i,j}$ is the $i$th Chern class in the $\mathbf{16}$ of the $j$th $SO(16)$ factor. Similarly for $E_7 \times SU(2) \times E_7 \times SU(2)$:\footnote{The term $X^{(E_7\times SU(2))^2}_8$ is related to the anomaly polynomial of the non-supersymmetric NS5 brane in this theory. It would be interesting to use anomaly inflow to shine light on its worldvolume degrees of freedom, along the lines of \cite{Basile:2023knk}.} 
\begin{align}
    X^{(E_7\times SU(2))^2}_4 &= \frac16(c_{2,\text{E1}}+c_{2,\text{E2}})+2 c_{2,1}+2 c_{2,2}+ p_1\,,\\
     X^{(E_7\times SU(2))^2}_8 &= \left(c_{2,1}-c_{2,2}\right) \left(c_{2,\text{E1}}-c_{2,\text{E2}}\right)\,,
\end{align}
where $c_{2,\text{Ei}}$ are $2$nd Chern classes in the $\mathbf{56}$ of the $i$th $E_7$ factor and $c_{2,j}$ are $2$nd Chern classes in the $\mathbf{2}$ of the $j$th $SU(2)$ factor. In contrast, the local anomalies of $E_{8,2}$ and $E_8 \times SO(16)$ cancel trivially, without the need for a Green-Schwarz term. This factorization \eqref{eq:factorize}, together with the Green--Schwarz mechanism, is reminiscent of the Ho\v{r}ava--Witten (HW) construction \cite{hw1,hw2} of supersymmetric $E_8 \times E_8$. In that context, the factorization is crucial for cancelling the anomalous variation of bulk topological terms in eleven-dimensional M-theory. In the present case, however, there is no direct analog of this mechanism. Moreover, it is not clear that such an analog would even be well-defined, since the 11d geometry cannot be taken to be macroscopic. Nevertheless, the similarity suggests that the two theories may have a similar M-theoretic picture.

Note that the exact choice of which pairs of gauge groups and matter we get are not explained by our picture.  In particular for the identifications to give distinct gauge groups they should act differently on the boundary $E_8\times E_8$ fields, as is motivated from the orbifold construction of \cite{Dixon:1986iz}, as was also used in \cite{Faraggi:2007tj}.

What we have found is a resolution to the no-go theorem for finding an M-theory lift of the non-supersymmetric heterotic E-type theories \cite{Montero:2025ayi}.  It was shown there that there cannot be an M-theory lift of anomaly inflow mechanism of Ho\v{r}ava-Witten for chiral theories, applicable to groups other than $E_8$ and $G_2$.  The resolution we have found here is that indeed the boundaries can be physically inseparable, thus not allowing a macroscopic separation of the boundaries to apply their argument to.

\subsubsection{Fate of $SO(16)\times SO(16)$ strings at strong coupling}
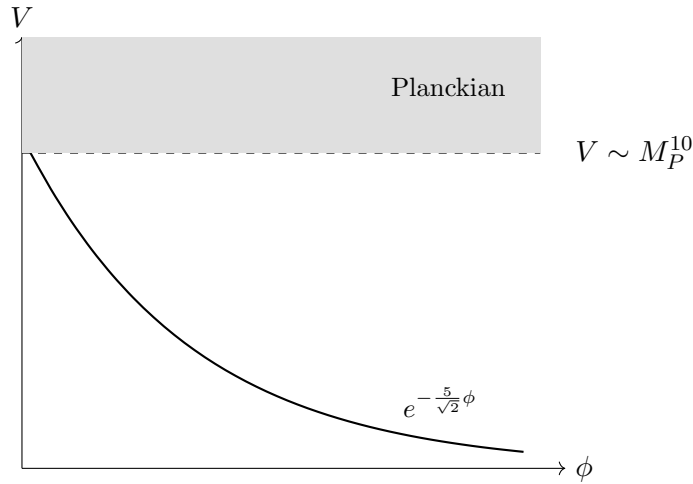
\begin{figure}[!h]
\centering
\begin{tikzpicture}[scale=1.1]

\draw[->] (0,0) -- (6.5,0) node[right] {$\phi$};
\draw[->] (0,0) -- (0,5.2) node[above] {$V$};

\draw[dashed] (0,3.8) -- (6.2,3.8);
\node[right] at (6.5,3.8) {$V \sim M_P^{10}$};

\draw[thick,domain=0.1:6,smooth] 
plot (\x,{4*exp(-0.5*\x)});

\fill[gray!25] (0,3.81) rectangle (6.2,5.2);

\node at (5.1,4.6) {\small Planckian};
\node at (5,0.8) {\small $e^{-\frac{5}{\sqrt{2}}\phi}$};

\end{tikzpicture}

\caption{
One-loop dilaton effective potential in the non-supersymmetric $SO(16)\times SO(16)$ theory. 
As the coupling increases, the potential should continue to rise and become Planckian before reaching the large-radius regime of M-theory to confine the singular geometry to sub-Planckian sizes. 
The shaded region denotes where the effective description is no longer reliable. We use the convention that large $\phi$ is weak coupling.
}
\label{fig:SO(16)xSO(16)}
\end{figure}

It is natural to ask what the fate of the non-supersymmetric tachyon-free $SO(16)\times SO(16)$ is at strong coupling.  The only scalar field in this theory is the dilaton for which a positive potential is generated at one loop \cite{Alvarez-Gaume:1986ghj} (the tachyon of 0A, related to the difference of the two radii, is projected out by the exchange symmetry requirement for the quotient and apparently no new ones arise between the DBRP boundaries). The coupling constant in this theory controls the size of the circle and since we do not expect the $\vee$ constructions to be macroscopically meaningful, they should be obstructed.  This suggests that the potential rises and becomes Planckian as we increase the coupling (and the radius) before the radius becomes much bigger than $O(1)$ in Planck units, see \autoref{fig:SO(16)xSO(16)}.  In other words this geometry is `confined' just as was proposed for the 0A case in \cite{Baykara:2026gem}.

\subsubsection{${E_8\times SO(16)}$}
The non-supersymmetric $E_8\times SO(16)$ theory is also consistent with DBRP, however there are some features that are different from the other two cases. First, the gauge groups are not identical even though they both can come from an $E_8$. It is reasonable that the identification of the boundaries can also be performed by breaking one $E_8\to SO(16)$ but not the other. The second feature is that there are no chiral fermions charged under both gauge fields. So there is a priori no reason for the branes to be stuck together, even though there is no obstruction for it either. So this theory can come from DBRP but it can also come from the usual $S^1/\bZ_2$ construction of $E_8\times E_8$ with an extra $\bZ_2$ action breaking the $E_8\rightarrow SO(16)$ on one of the walls. We leave the determination of which of these two scenarios is correct to future research.

\section{F-theory $\bZ_2$ quotients on $(S^1\vee S^1)\times S^1$}\label{sec:F-theory}
In this section we consider $\bZ_2$ quotients of $(S^1\vee S^1)\times S^1$ and we argue that we can find representations of all three of the 0B orientifolds (including the non-tachyonic $U(32)$ theory of Sagnotti \cite{Sagnotti:1995ga}) as well as non-supersymmetric heterotic strings of the D-type.  The three orientifolds of 0B include reflection on the stand-alone $S^1$, and they differ by whether we reflect one or two circles of $S^1 \vee S^1$ or exchange them, i.e., in the notation of 0A as $(P^\pm, P^+P^-,E)$. We will discuss each one separately, starting with $0B/\Omega$ in \autoref{sec:0BOmega}, then the Sagnotti string $0B/\Omega (-1)^{G_R}$ in \autoref{sec:sagnotti} and finally, we will consider $0B/\Omega (-1)^{F_R}$ in \autoref{sec:0BomegaFR}. The non-supersymmetric heterotic ones are obtained by putting the E-type M-theory constructions we mentioned on a circle with holonomy and taking the volume to zero size, leading to an F-theory picture where we exchange the $S^1\overset{\overset{E}\leftrightarrow}\vee S^1$ without reflecting the stand-alone circle. We will discuss them in \autoref{sec:D-type}.

\subsection{0B orientifold: $(S^1\vee S^1) \times ({ S}^1\updownarrow$ )}\label{sec:0BOmega}

The Type 0B string can be thought of as the orbifold of the Type IIB string by $(-1)^F$. Its torus partition function reads
\begin{equation}
    \mathcal{T} \sim O_8 \bar O_8 + V_8 \bar V_8 + S_8 \bar S_8 + C_8 \bar C_8\,,
\end{equation}
with the last two terms counting the untwisted and twisted RR sectors. As before, we denote the untwisted RR 0-, 2- and 4-forms as $\chi,B,D$, whereas $\chi',B',D'$ are the twisted ones and define the combinations
\begin{equation}
    \chi^\pm = \chi \pm \chi'\,, ~~~~~ B^\pm = B \pm B', ~~~~~~ D^\pm = D \pm D'\,. 
\end{equation}
It is clear that this orbifold preserves the parity symmetry $\Omega$ of the parent Type IIB string, which acts separately on each RR sector preserving only the 2-forms $B,B'$. Its action on the NSNS sector is as usual, preserving the metric $g$, the dilaton $\phi$ and the tachyon $T_{0B}$. 

Through the standard computation of the Klein bottle amplitude we see that this orientifold has a vanishing RR tadpole but a doubled NSNS tadpole, relative to the orientifold $IIB/\Omega$ \cite{Bergman:1997rf}. Again, we choose to cancel the NSNS tadpole by introducing in total 64 $D9$-branes. Since these D-branes have non-vanishing RR-charge, they must be introduced in brane-anti-brane pairs, but these pairs can be either of the $D9^+$-type or $D9^-$-type. For $n$ pairs of the plus type, the gauge symmetry group reads
\begin{equation}
    G = (SO(n)\times SO(32-n))^2\,.
\end{equation}
The rest of the open string sector comprises massless Majorana-Weyl fermions in the representations
\begin{equation}
    (\Box,\Box,1,1)_+\,, ~~~~~ (\Box,1,1,\Box)_-\,, ~~~~~ (1,1,\Box,\Box)_+\,, ~~~~~ (1,\Box,\Box,1)_-\,,
\end{equation}
where the subscript denotes the spacetime chirality, as well as tachyons in the representations
\begin{equation}
    (\Box,1,\Box,1)\,, ~~~~~ (1,\Box,1,\Box)\,.
\end{equation}

\begin{figure}
    \centering
    \begin{tikzpicture}[>=Latex, line cap=round, line join=round, scale=1.2]

\draw[thick] (-3,0) circle (0.8);
\draw[thick] (-1.4,0) circle (0.8);

\node at (-0.3,0) {$\times$};

\draw[thick] (0.8,0) circle (0.8);

\draw[very thick,<->] (0.8,0.62) -- (0.8,-0.62);

\node at (-2.2,-1.5) {$S^1 \vee S^1$};
\node at (0.8,-1.5) {$S^1$};

\draw[thick,->] (2,0) -- (3,0);

\begin{scope}[shift={(4.,0)}]
    \draw[thick] (-.6,1.2) arc[start angle=180,end angle=360,x radius=0.6,y radius=0.22];
    \draw[thick] (-.6,1.2) arc[start angle=180,end angle=0,x radius=0.6,y radius=0.22];

    \draw[thick] (-.6,-1.2) arc[start angle=180,end angle=360,x radius=0.6,y radius=0.22];
    \draw[dashed] (-.6,-1.2) arc[start angle=180,end angle=0,x radius=0.6,y radius=0.22];

    \draw[thick] (-0.6,1.2) -- (-0.6,-1.2);
    \draw[thick] (0.6,1.2) -- (0.6,-1.2);
\end{scope}

\begin{scope}[shift={(5.2,0)}]
    \draw[thick] (-.6,1.2) arc[start angle=180,end angle=360,x radius=0.6,y radius=0.22];
    \draw[thick] (-.6,1.2) arc[start angle=180,end angle=0,x radius=0.6,y radius=0.22];

    \draw[thick] (-.6,-1.2) arc[start angle=180,end angle=360,x radius=0.6,y radius=0.22];
    \draw[dashed] (-.6,-1.2) arc[start angle=180,end angle=0,x radius=0.6,y radius=0.22];

    \draw[thick] (0.6,1.2) -- (0.6,-1.2);
\end{scope}

\draw[very thick] (4.6,1.2) -- (4.6,-1.2);

\node at (4.6,-1.75) {$(S^1\vee S^1)\times (S^1\updownarrow)$};

\end{tikzpicture}
    \caption{F-theory geometry corresponding to $0B/\Omega$. The space $(S^1\vee S^1)\times S^1$ is quotiented by a reflection on the stand-alone $S^1$, resulting in two cylinders stuck together along an interval.}
    \label{fig:0BOmega}
\end{figure}
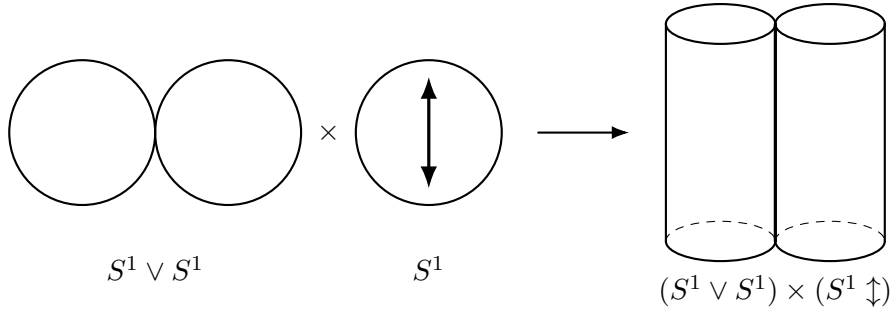

Compactifying this theory on a circle and performing T-duality, we obtain an orientifold of the Type 0A string by $\Omega R$, with $R$ the usual reflection of the dual $S^1$. We may wonder if, since this is the Type 0A theory, we should identify $\Omega$ with $\Omega^-_{0A}$. This would imply that the M-theory picture involves a reflection of $S^{1-}$ together with the stand-alone $S^1$. However, we can obtain this Type 0A orientifold by quotienting the Type I' theory (orientifold of Type IIA by $\Omega R$) by $(-1)^{F}$. The M-theory uplift of Type I' has the geometry of a cylinder $S^1\times I$, where the action of $\Omega$ corresponds to parity $\hat \Omega$ of the M2 brane. It is thus more natural that the M-theory geometry of the 0A orientifold, and thus the F-theory geometry of its 10d 0B counterpart, is in fact $(S^1\vee S^1)\times I$, see \autoref{fig:0BOmega}. In summary, the geometric picture is given by M-theory on $(S^1\vee S^1) \times S^1$ quotiented by 
\begin{equation}\label{0Blift1}
    \begin{split}
        \Omega &\to \hat \Omega: ~~~~~ C \mapsto -C\,,\\\\
        R &\to R: ~~~~~ (S^1\vee S^1)\times  S^1 \to (S^1\vee S^1)\times (S^1 \updownarrow) \,,\\
        &~~~~~~~~~~~~~~~ x^9 \mapsto -x^9\,.
    \end{split}
\end{equation}

Let us first explain how the closed string spectrum is recovered in this picture. Although this case is rather simple, it will make our analysis of the remaining three orientifolds easier. We obtain the M-theory fields on $(S^1\vee S^1 )\times S^1$ by reducing the $(S^1\vee S^1)$ fields on $S^1$. From the 1-forms and 3-forms we get in 9d
\begin{equation}
 A^+\,,~ A^-\,,~ C^+\,, ~ C^-\,, ~ \chi^+\,, ~ \chi^-\,, ~ B^+\,, ~ B^-\,.
\end{equation}
From the metric $g$ and the B-field $B_{NS}$ we get two extra vectors $A_{NS}$ and $A_{NS}'$, and we also have the dilaton $\phi$ and the tachyon $T_{0A}$. The reflection $R$ acts as $-1$ on the reduced fields $\chi^\pm$, $B^\pm$, $A_{NS}$ and $A'_{NS}$, whereas $\hat \Omega$ acts as $-1$ on $C^\pm$, $B^\pm$, $B_{NS}$ and $A_{NS}'$. Hence after quotienting by $\hat \Omega R$ we have the fields $A^\pm$, $B^\pm$, $A_{NS}'$, $g$, $T_{0A}$ and $\phi$ in 9d. T-dualizing on $S^1$ and going to the 10d 0B picture, $A^\pm$ and $A_{NS}'$ get absorbed in the 10d $B^\pm, g_{\mu\nu}$ and we are left with the closed string spectrum of $0B/\Omega$.  

To explain the $D9^\pm$ branes in this setup, we recall that in the M-theory realization we first put the Type I theory on a circle with a holonomy which breaks the $SO(32)\rightarrow SO(16)\times SO(16)$, corresponding to equal amounts of D8 branes on the Type I' boundaries, and the D8 branes lift to M9 branes of M-theory (wrapping the 11th direction).  Similarly here we will assume the same for $D9^\pm$ and their anti-branes.  Namely in going to 9 dimensions we turn on suitable holonomy corresponding to putting an equal number of $M9^\pm$ and $\overline{ M9^\pm}$ (wrapping the corresponding circles) at the two ends of the interval, namely $\frac{n}{2}$ ($M9^+,\overline{M9^+}$) and $\frac{32-n}{2}$($M9^-,\overline{M9^-}$) at each boundary. It is not difficult to explain the expected open string matter content discussed above for considerations to the discussions in \cite{Baykara:2026gem}.

We will consider the problem of tachyon condensation in more detail when we connect this theory (with $n = 0$) to the bosonic string following Bergman and Gaberdiel \cite{Bergman:1997rf}.

\subsection{Tachyon-free $U(32)$ orientifold: $(S^1\overset{\overset{Q}\leftrightarrow }\vee S^1)\times (S^1\updownarrow)$}\label{sec:sagnotti}

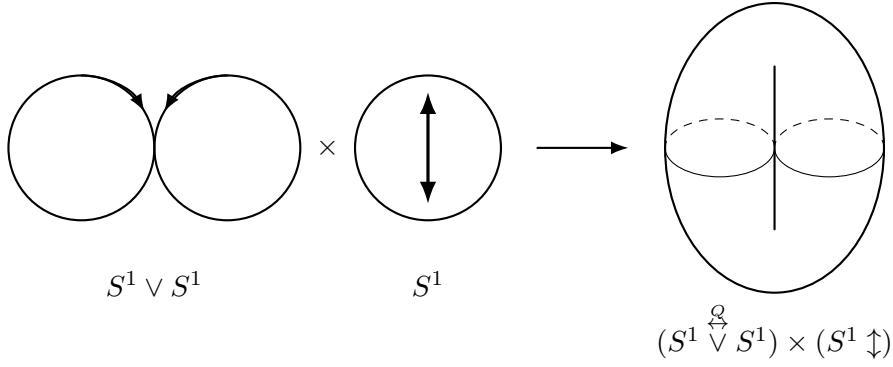
\begin{figure}
    \centering
    \begin{tikzpicture}[>=Latex, line cap=round, line join=round, scale=1.2]

\draw[thick] (-3,0) circle (0.8);
\draw[thick] (-1.4,0) circle (0.8);
\draw[->,thick] (-3,0.8) arc[start angle=90,end angle=30,radius=0.8];

\draw[->,thick] (-1.4,0.8) arc[start angle=90,end angle=150,radius=0.8];

\node at (-0.3,0) {$\times$};

\draw[thick] (0.8,0) circle (0.8);

\draw[very thick,<->] (0.8,0.62) -- (0.8,-0.62);

\node at (-2.2,-1.5) {$S^1 \vee S^1$};
\node at (0.8,-1.5) {$S^1$};

\draw[thick,->] (2,0) -- (3,0);

\begin{scope}[shift={(4.6,0)}]

\draw[thick] (0,0) ellipse (1.2 and 1.6);

\draw[thick] (0,0.9) -- (0,-0.9);

\draw[dashed] (0,0) arc[start angle=0,end angle=180,x radius=0.6,y radius=0.32];
\draw (0,0) arc[start angle=0,end angle=-180,x radius=0.6,y radius=0.32];

\draw[dashed] (1.2,0) arc[start angle=0,end angle=180,x radius=0.6,y radius=0.32];
\draw (1.2,0) arc[start angle=0,end angle=-180,x radius=0.6,y radius=0.32];

\end{scope}

\node at (4.6,-2) {$(S^1\overset{\overset{Q}\leftrightarrow}\vee S^1)\times (S^1\updownarrow)$};

\end{tikzpicture}
    \caption{F-theory geometry of $0B/\Omega(-1)^{G_R}$ obtained by quantum symmetry $Q$ on the $S^1\vee S^1$ and reflection on the stand-alone $S^1$. Over an interval, the generic fiber has $S^1\vee S^1$ and at the endpoints the fiber is $S^1$. Note that quantum symmetry $Q$ corresponds to shifting in both the CRP and CRP$'$ resolutions, so the action can not be faithfully shown on a single $S^1\vee S^1$. The figure above shows the identification for the CRP$'$ resolution which corresponds to the under-over crossing resolution of $S^1\vee S^1$.}
    \label{fig:0BQ}
\end{figure}

Now we consider the more well known orientifold of the Type 0B string \cite{Sagnotti:1995ga} given by $\Omega (-1)^{G_R}$ (where worldsheet fermion number $(-1)^{G_R}=Q$ is the quantum symmetry of 0B), which has gauge group $U(32)$ and is free from tachyons. Compactifying on $S^1$ and T-dualizing we obtain Type 0A on $S^1$ quotiented by $\Omega R(-1)^{G_R}$.  Using the lifts of $\Omega$ and $R$ as discussed in eq. \eqref{0Blift1}, and using the fact that $(-1)^{G_R}=Q$  is the quantum symmetry exchange of the circles (corresponding to shift on the covering circle) we obtain the F-theory picture
\begin{equation}
\begin{split}
    (S^1\overset{\overset{Q}\leftrightarrow }\vee S^1)\times (S^1\updownarrow)
\end{split}
\end{equation}
Since $Q$ acts as a translation on $S^1\vee S^1$, the geometric action given by $R Q$ is free, thus there are no fixed points in the quotient, see \autoref{fig:0BQ}.

The closed string spectrum of the 0B orientifold is obtained as follows. The quantum symmetry $(-1)^{G_R}$ acts as $-1$ on the twisted fields of $0B = IIB/(-1)^F$, i.e. the tachyon $T_{0B}$ and the RR fields $\chi', B', D'$, thus the combined action $\Omega (-1)^{G_R}$ projects out $B_{NS}$, $T_{0B}$ and $B'$. We then have the fields
\begin{equation}
    g\,,~ \phi\,,~ \chi'\,,~ B\,, ~ D'\,. 
\end{equation}
To see this from the M-theory point of view, we complement the analysis of the previous subsection by including the action of $Q$. For any pair of fields $\phi^\pm$, $Q$ acts as $+1$ ($-1$) on the (anti)diagonal combination $\phi^+ + \phi^-$ ($\phi^+ - \phi^-$). Moreover, since it identifies the two circles in $S^1\vee S^1$, it projects out the tachyon $T_{0A}$. The overall quotient by $\hat \Omega R Q$ thus preserves exactly the fields corresponding to $0B/\Omega$ compactified on $S^1$. 

This 0B orientifold has the particular property of having a vanishing NS-NS tadpole in the Klein bottle. In other words, there is no tensionful object that we can identify with a standard O-plane. This is consistent with our picture of a geometry without fixed points. 

The O9 plane has vanishing NSNS coupling but non-vanishing RR charge, and the only way to cancel the RR tadpole is to introduce 32 $D9^+$- and $D9^-$-branes, and possibly extra brane-antibrane pairs. The symmetry $\Omega (-1)^{G_R}$ trades the Chan-Paton degrees of freedom of the $D9^+$- and $D9^-$-brane stacks, thus projecting $U(32)\times U(32) \to U(32)$.  Moreover we find a chiral fermion in the rank 2 antisymmetric representation and its conjugate with the same chirality.

As for the open string sector, it is natural to propose that the $D9$-branes are associated to M9-branes reduced on the $S^{1\pm}$, which become identified, leading to a single $U(32)$.

\subsection{0B orientifold with $(-1)^{F_R}$: $\big[(S^1\vee S^1) \times S^1\big]\updownarrow $}\label{sec:0BomegaFR}
We now consider the last orientifold $0B/\Omega(-1)^{F_R}$ \cite{Bergman:1997rf}. The perturbative spectrum consists of the metric $g_{\mu\nu}$, the axio-dilaton $\tau$, the complexified tachyon
\begin{align}
    z = T + i\chi'
\end{align}
of 0B, and $D^\pm$ forming an unconstrained 4-form. There is no RR tadpole to cancel, so there is no need for space-filling D9 branes and as a consequence there is no gauge field unlike the previous 0B orientifold case.

Note that $(-1)^{F_R}$ acts as a $-1$ on all RR fields and trivially on all other fields. Therefore we can identify $(-1)^{F_R}$ as the reflection on the two circles of $S^1\vee S^1$
\begin{align}
(-1)^{F_R}:\theta^\pm &\mapsto -\theta^\pm.
\end{align}
Combining with the action of $\Omega$ in 0B, we find that the zero volume limit of M-theory that corresponds to the 10d orientifold is
\begin{align}\label{eq:0BOmegaFL}
    0B/\Omega(-1)^{F_R} = \text{M-theory on }\big[(S^1 \vee  S^1) \times  S^1\big]\updownarrow \Big\vert_{V=0}.
\end{align}

\begin{figure}
    \centering
    \begin{tikzpicture}[>=Latex, line cap=round, line join=round, scale=1.2]

\draw[thick] (-3,0) circle (0.8);
\draw[thick] (-1.4,0) circle (0.8);
\draw[very thick,<->] (-3,0.62) -- (-3,-0.62);
\draw[very thick,<->] (-1.4,0.62) -- (-1.4,-0.62);

\node at (-0.3,0) {$\times$};

\draw[thick] (0.8,0) circle (0.8);

\draw[very thick,<->] (0.8,0.62) -- (0.8,-0.62);

\node at (-2.2,-1.5) {$S^1 \vee S^1$};
\node at (0.8,-1.5) {$S^1$};

\draw[thick,->] (2,0) -- (3,0);

\begin{scope}[shift={(4.0,0)}]

\coordinate (A) at (-0.8,1.0);
\coordinate (B) at (0.8,1.0);
\coordinate (C) at (0.8,-1.0);
\coordinate (D) at (-0.8,-1.0);

\draw[thick] (A) .. controls (0,0.6) .. (B);
\draw[thick] (C) .. controls (0,-0.6) .. (D);
\draw[thick] (D) .. controls (-0.4,0) .. (A);

\end{scope}

\begin{scope}[shift={(5.6,0)}]

\coordinate (A) at (-0.8,1.0);
\coordinate (B) at (0.8,1.0);
\coordinate (C) at (0.8,-1.0);
\coordinate (D) at (-0.8,-1.0);

\draw[thick] (A) .. controls (0,0.6) .. (B);
\draw[thick] (B) .. controls (0.4,0) .. (C);
\draw[thick] (C) .. controls (0,-0.6) .. (D);

\end{scope}

\draw[very thick] (4.8,1.0) -- (4.8,-1.0);

\begin{scope}[shift={(3.58,0.59)}, rotate=42.62]

    \draw[dashed] plot[domain=0:180,smooth,variable=\t]
        ({0.24*cos(\t)}, {0.075*sin(\t)});

    \draw plot[domain=180:360,smooth,variable=\t]
        ({0.24*cos(\t)}, {0.075*sin(\t)});
\end{scope}
\draw[dashed] (4.8,0) arc[start angle=0,end angle=180,x radius=0.65,y radius=0.17];
\draw (4.8,0) arc[start angle=0,end angle=-180,x radius=0.65,y radius=0.17];

\draw[dashed] (6.1,0) arc[start angle=0,end angle=180,x radius=0.65,y radius=0.17];
\draw (6.1,0) arc[start angle=0,end angle=-180,x radius=0.65,y radius=0.17];

\node at (4.8,-1.5) {$[(S^1\vee S^1)\times S^1]\updownarrow$};

\end{tikzpicture}
    \caption{F-theory geometry of $0B/\Omega(-1)^{F_R}$ obtained by quotienting by reflecting all circles. As a result, we get two pillowcase orbifolds $T^2/\bZ_2$ stuck together along an edge.}
    \label{fig:0BFL}
\end{figure}
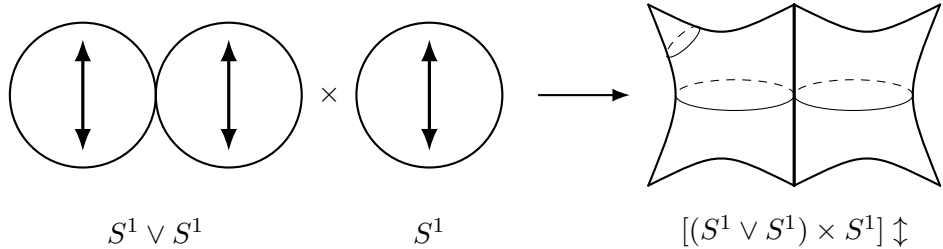

The action on the geometry gives another explanation for which 0B fields are projected out and which are kept. The axiodilatons $\tau_+$ and $\tau_-$ are kept as they correspond to the two complex parameters associated with the geometry. The axiodilatons recombine to give $\tau$ and $z$.\footnote{Similar to the 0B proposal in \cite{Baykara:2026gem} it is natural to view this as the moduli space $\tau$ of $T^2/\bZ_2$ with a point $z$ on it.} Additionally, $D^\pm$ is kept since on the M-theory side it corresponds to 3-forms $C^\pm$ on spacetime for which the effect of circle reflections and orientifolding cancel out. The fields that are projected out correspond to the $A^\pm$ on spacetime and $C^\pm$ with two legs on the compactification, which on the 0B side correspond to $B_{RR}^\pm$. See \autoref{fig:0BFL} for the geometry.

Differently from the other cases, this orientifold can also be obtained directly from IIB as
\begin{align}
    IIB/\Omega(-1)^{F_R} = 0B/\Omega(-1)^{F_R}.
\end{align}
To see this, note that 0B is an orbifold of IIB
\begin{align}
    0B = IIB/(-1)^F
\end{align}
and that $(-1)^F = (\Omega(-1)^{F_R})^2$. So the IIB orientifold has order $\bZ_4$, with the second twisted sector corresponding to 0B. 

The existence of the IIB perspective is also hinted from F-theory perspective by reversing the order of $\vee , \bZ_2$ operations. 
In particular if we do the $\bZ_2$ quotient first, the $\vee$ operation is redundant as the point has been already identified by the involution (if we ignore the action on the stand-alone circle).  This suggests that we start with IIB given by F-theory on $T^2$, and lift the action $\Omega(-1)^{F_R}$ as a geometric $\bZ_2$ reflection action on $T^2$, which squares to $(-1)^F$ and so becomes $\bZ_4$ when acting on the full spectrum with fermions.\footnote{We will refer to the compactification as $T^2/\bZ_2$ to emphasize the geometric action, but it should be kept in mind that the action lifts to $\bZ_4$. This orientifold realized as a geometric M-theory compactification is similar in spirit to the one recently discussed in \cite{bcd}, where however there the $\bZ_2$ orbifold action was accompanied by a freely-acting shift in an additional circle.} In other words this suggests that M-theory on $T^2/\bZ_2$ in the zero volume limit also produces the 10d orientifold\footnote{Just as in the 0A case it is natural to believe that the M-theory version starts with a $T^2$ with a non-trivial even spin structure and then quotienting it by $\bZ_2$.}
\begin{align}\label{eq:T2modZ2}
0B/\Omega(-1)^{F_R} = \text{M-theory on }T^2/\bZ_2 \Big\vert_{V_{T^2}=0}.
\end{align}

Given \eqref{eq:0BOmegaFL} and \eqref{eq:T2modZ2} as two different $0B/\Omega(-1)^{F_R}$ limits, one might wonder if they are equivalent at finite volume in 9d as well. This cannot be the case for the same reasons that the two descriptions of 0B are equivalent only in 10d but not in 9d as explained in detail in section 5 of \cite{Baykara:2026gem}. In particular, in 9d the 0B descriptions differ by a choice of holonomy on $S^1$ for the 0B side. This means they are different theories in 9d but arise from the same 10d theory with distinct twist actions on the circle.

\subsection{D-type heterotic strings: $(S^1\overset{\overset{E}\leftrightarrow}\vee S^1) \times S^1_\gamma$} \label{sec:D-type}
The geometrization of the D-type non-supersymmetric strings is analogous to the geometrization of the  supersymmetric $SO(32)$ heterotic using the $E_8\times E_8$ heterotic string. 

For review, putting the $E_8\times E_8$ heterotic string on a circle and turning on holonomy $\gamma$ is T-dual to the $SO(32)$ heterotic string. So the 10d $SO(32)$ heterotic string can also be viewed from the Ho\v{r}ava-Witten M-theory interpretation as M-theory on a cylinder at zero volume
\begin{align}
    \text{M-theory on }I\times S^1_\gamma \big\vert_{V=0}= E_8\times E_8 \text{ Het on } S^1_\gamma\Big\vert_{R=0} = SO(32) \text{ Het}.
\end{align}
Note that at finite radius, the duality is in 9d, where the gauge group is $SO(16)\times SO(16)$ throughout with each $SO(16)$ factor coming from one of the $E_8$s. Only at zero volume the two factors recombine to give $SO(32)$.

We now extend this pattern to the present case. The D-type non-supersymmetric heterotic strings are the analogous duals of the E-type heterotic strings. In particular, consider the E-type heterotic string construction proposed in \autoref{sec:e-type-het} on a circle with holonomy $\gamma$
\begin{align}
    S^1\overset{\leftrightarrow}{\vee} S^1 \times S^1_\gamma.
\end{align}
By various choices of the starting point of the E-type string and the holonomy, all the D-type heterotic strings can be obtained in 10d,
\begin{align}
    SO(32), SO(24)\times SO(8), SU(16)\times U(1).
\end{align}
Similar to the supersymmetric case, in the 9d interpolation the gauge groups of E-type heterotic strings are broken and then recombine at the zero volume radius giving the gauge groups of D-type strings. 

\section{Dualities with Bosonic Strings}
\label{sec:dualitiesbosonic}

Having discussed $\bZ_2$
quotients of M-theory and F-theory descriptions of 0A and 0B and argued that they include all the non-supersymmetric 10d heterotic strings as well as 0A and 0B orientifolds, we now turn to dualities between superstring theories and bosonic string theories. The first is the duality proposed in \cite{Bergman:1997rf}, and the second was proposed in \cite{Dudas:2001wd}. We discuss the first one in some detail, and propose resolution of some of the mismatch of fields demanded by duality as well as provide additions tests for the duality.  The second duality also has field mismatches and we also resolve this by a similar method.

\subsection{Bergman-Gaberdiel Duality}\label{sec:BG}

We now turn to the duality conjectured in \cite{Bergman:1997rf}, which relates a Type 0B orientifold (a particular case of those discussed in Section~\ref{sec:0BOmega}) to a bosonic string compactified on a $T^{16}$ torus. We begin in Section~\ref{sec:BGsetup} with a review of the proposed duality. We then summarize the evidence presented in \cite{Bergman:1997rf} in support of this correspondence, highlighting several outstanding puzzles. In Section~\ref{sec:resolutionBG}, we propose a resolution to these issues, and finally provide further evidence for the duality in \autoref{sec:BGfurther}.

\subsubsection{Statement of the duality}\label{sec:BGsetup}
One side of the duality corresponds to the Type 0B orientifold described in \autoref{sec:0BOmega} with the choice of $n=0$ background corresponding to $32(D9^-,\overline{D9^-})$ branes. For this special choice, there are no massless fermions, and there is only one tachyon in the bifundamental $(\mathbf{32},\mathbf{32})$. Summarizing the discussion in \ref{sec:0BOmega}, we are left with a theory of unoriented open and closed strings with gauge group $SO(32)\times SO(32)$ whose massless and tachyonic spectrum is shown on the left column in \autoref{fig:tablespectrum}.

\begin{figure}
    \centering
    \[
\begin{array}{c|c}
0B/\Omega & \text{Bosonic String on } T^{16}_{SO(32)_L\ \otimes SO(32)_R} \\
\hline
g_{\mu\nu},\phi & g_{\mu\nu},\phi \\
A_{\mu} & A_{\mu}\\
B_{\mu\nu}^\pm & B_{\mu\nu}\\
(\mathbf{1},\mathbf{1})~T & (\mathbf{1},\mathbf{1})~T \\
(\square,\square)~T & (\square,\square)~T \\
 & (\,\vcenter{\hbox{\ydiagram{1,1}}}\,,\vcenter{\hbox{\ydiagram{1,1}}}\,)\supset~\text{Narain moduli}
\end{array}
\]
    \caption{Massless and tachyonic spectrum of the two proposed dual theories. There is a mismatch by an extra 2-form field on the $0B/\Omega$ side and (tree-level) massless scalars in the $(\,\vcenter{\hbox{\ydiagram{1,1}}},\vcenter{\hbox{\ydiagram{1,1}}}\,)$ representation, with Cartan directions corresponding to Narain moduli.}
    \label{fig:tablespectrum}
\end{figure}

The other side of the duality is given by a bosonic string theory compactified on a $T^{16} $ torus. In 26 dimensions, bosonic string theory has a massless spectrum consisting of the metric $G_{\mu\nu}$, the dilaton $\phi$, and the Kalb--Ramond $B$-field $B_{\mu\nu}$. It contains no perturbative fermions and includes a tachyon $T$ with $\alpha' M^2 = -2$. We now proceed to compactifying this theory on a 16-dimensional torus $T^{16}$. The bosonic string on $T^{16}$ has a point in moduli space where the gauge group is locally $SO(32)^2$ with a bifundamental tachyon. It corresponds to the Narain lattice\footnote{There is another Narain lattice that leads to a gauge group which is locally $SO(32)^2$: $\Gamma_{16,16} = D_{16}^+ \oplus D_{16}^+\,$, where $D_{16}^+$ is the weight lattice of $Spin(32)/\mathbb{Z}_2$. The gauge group is globally $Spin(32)^2/\mathbb{Z}_2^2$, and there are no charged tachyons at all.  This compactification will not be used in this paper.}
\begin{equation}
    \Gamma_{16,16} = [D_{16}\oplus D_{16}]\,,
\end{equation}
where by this we mean the lattice where left and right momenta are in the weight lattice of $D_{16}$ with the restriction that the difference of the left and right momenta belong to the root lattice. In other words, 
\begin{equation}
    [D_{16}\oplus D_{16}]
    \;=\;
    \left\{
    (p_L,p_R)\in D_{16}^{*}\oplus D_{16}^{*}
    \;\middle|\;
    p_L-p_R\in D_{16}
    \right\},
\end{equation}
where \(D_{16}\) is the root lattice of \(\mathfrak{so}(32)\), and
\(D_{16}^{*}/D_{16}=\{o,v,s,c\}\) denotes the four conjugacy classes
(root, vector, spinor, conjugate-spinor).  Thus the lattice sum contains
precisely the sectors
\begin{equation}
    (o,o)\oplus (v,v)\oplus (s,s)\oplus (c,c)\, .
\end{equation}
At a generic point in the toroidal moduli space the compactification on
\(T^{16}\) gives an abelian gauge algebra $   \mathfrak{u}(1)^{16}_{L}\oplus \mathfrak{u}(1)^{16}_{R}$, whose gauge bosons come from \(G_{\mu I}\) and \(B_{\mu I}\),
with \(I=1,\dots,16\).  At the special point
\(\Gamma_{16,16}=[D_{16}\oplus D_{16}]\), additional lattice states become
massless.  These are the states with either $    p_L^2=2,\; p_R=0$ or $    p_R^2=2, \;p_L=0.$
Since the norm-two vectors in \(D_{16}\) are precisely the roots of
\(\mathfrak{so}(32)\), the gauge algebra is enhanced to
\begin{equation}
    \mathfrak{so}(32)_{L}\oplus \mathfrak{so}(32)_{R}.
\end{equation} There are additional scalar fields arising from the same lattice sectors.
These scalars transform in the bi-adjoint representation. Furthermore, the \((v,v)\) sector contains
states with
\begin{equation}
    p_L^2=p_R^2=1,\qquad N_L=N_R=0,
\end{equation}
so that
\begin{equation}
    M^2
    \;=\;
    p_L^2+2(N_L-1)
    \;=\;
    p_R^2+2(N_R-1)
    \;=\;
    -1.
\end{equation}
These states transform in the \((\mathbf{32},\mathbf{32})\) of
\(SO(32)_L\times SO(32)_R\), and are tachyonic in ten dimensions. Putting all of this together we obtain the massless and tachyonic spectra listed on the right-hand column of 
\autoref{fig:tablespectrum}.

\subsubsection{Evidence \& Puzzles}
We now turn to the evidence in favour of this duality, and outline a few outstanding puzzles first pointed out in \cite{Bergman:1997rf}. 

\paragraph{A solitonic bosonic string}
The fantastic realization of \cite{Bergman:1997rf} is that one of the two D1-brane in the Type 0B orientifold carries exactly the same degrees of freedom as the fundamental bosonic string in the dual picture! We now review this observation, starting with the brane content of the Type 0B orientifold. In the parent Type 0B theory the R-R sector is doubled, and correspondingly the D-brane spectrum is doubled as well: for each allowed value of \(p\) there are two elementary branes, which we denote by \(Dp^{+}\) and \(Dp^{-}\). As reviewed in \autoref{sec:0BOmega}, the orientifold projects out both the RR axions and 4-forms. The corresponding branes are projected out as well.\footnote{We will discuss the torsion charged D-branes of the Type 0B orientifold in the context of this duality in \autoref{sec:BGfurther}.} The two $D1^\pm$ branes which couple electrically to $B^{\pm}$ remain in the spectrum. Let us study the massless excitations on these two 1-branes. These are described by two different types of open strings: those beginning and ending on the 1-brane, and those beginning and ending on the background 9-branes \cite{Polchinski:1995df}. 

As explained in \cite{Bergman:1997rf}, open strings stretching between the same D1-brane give rise to eight left-moving and eight right-moving scalars on the D1 worldsheet. For strings stretching between a D1-brane and a D9-brane, the spectrum depends on the choice of $\pm$ variants. Without loss of generality, we take the background 9-branes to all be of $D9^{-}$  and $\overline{D9^-}$ type. In this background, open strings stretching between a $D1^{-}$ and a $D9^{-},\overline{D9^-}$'s do not produce massless modes on the $D1^{-}$ worldsheet: the R–R boundary state contribution vanishes, while the NS–NS contribution is massive. Putting everything together, the worldsheet theory of the $D1^{-}$ contains only eight left- and eight right-moving scalars, parametrizing the embedding of the string in ten-dimensional spacetime.

Instead consider the $D1^+$ string.  The open strings stretched between a $D1^{+}$ and background of $D9^{-},\overline{D9^-}$'s lead to massless open string states in the twisted R sector: each 9-brane contributes a single massless fermion of definite chirality, while each anti-9-brane contributes a fermion of the opposite chirality. Altogether, the spectrum consists of eight left- and eight right-moving scalars, together with 32 left- and 32 right-moving fermions transforming in the $(\mathbf{32},\mathbf{1})$ and $(\mathbf{1},\mathbf{32})$ of $SO(32)\times SO(32)$, respectively. This precisely matches the worldsheet content of the dual bosonic string on the $T^{16}$ torus described above which corresponds to its bosonization.

One can repeat the same analysis for the other $SO(n)^2 \times SO(32-n)^2$ orientifolds, but none of them admits a D1-brane with the correct degrees of freedom to correspond to any known critical string theory. Perhaps relatedly, all of these other orientifolds contain massless chiral fermions, which cannot be made massive, making it hard to envision how a purely bosonic theory could emerge as a dual.

This suggests an appealing picture in which the $D1^+$ string of $0B/\Omega$ becomes light at strong coupling and is identified with the fundamental string of the dual bosonic theory. Moreover, we must understand why the $D1^{-}$ is absent in the bosonic dual. We will propose a resolution to this puzzle in \autoref{sec:resolutionBG}.

\paragraph{Matching the massless and tachyonic spectra}

Reading off \autoref{fig:tablespectrum}, we find that the massless spectra of the two theories are nearly identical. Both contain a metric, a dilaton, and a $B$-field, along with vector bosons transforming in the adjoint of $SO(32)\times SO(32)$. On the 0B side, this matching occurs only for the special choice $n=0$, for which the spectrum is free of massless chiral fermions. Such fermions could not admit a dual description in a bosonic theory, as there is no mechanism to give them a mass.

Both theories also have the same tachyonic spectrum. On the bosonic side, the singlet tachyon comes from the ground state $\ket{0}\otimes \ket{0}$ and has a mass $\alpha'_B M^2 =- 4$, whilst the $(\mathbf{32},\mathbf{32})$ tachyon comes from states of the form $    \psi^a_{-1/2} \tilde{\psi} ^b _ {-1/2} \ket{0} \otimes \ket{0}$ with mass $\alpha_B' M^2 = -2$. On the 0B side, the ground state singlet tachyon has a mass $\alpha'_0 M^2 = -2$ whilst the charged tachyon comes from the open string sector and has a mass $\alpha' _ 0M^2 = -1/2$. In the absence of supersymmetry, masses are expected to receive corrections, so it is not problematic that they do not match; instead, we focus on representations of the gauge group. In particular, the presence of $(\mathbf{32},\mathbf{32})$ states on both sides provides evidence for a duality. The existence of two singlet tachyons is not, by itself, evidence for a duality, but they will play a crucial role in Section~\ref{sec:resolutionBG}.  

The massless and tachyonic spectrum on both sides of the theory is almost identical except for two puzzles in this matching. The first is to explain the absence of the Narain moduli on the Type 0B orientifold side. The second is to account for the additional $B$-field present on the 0B side. We will discuss possible resolutions for both puzzles in the next section.

\subsubsection{Resolution of the mismatch puzzles}\label{sec:resolutionBG}
We now turn to our proposed resolutions for the puzzles outlined in the previous section. 
\paragraph{Massing up Narain moduli}

The first puzzle is to understand why the Narain moduli appear in the massless spectrum of the bosonic theory, but are absent in the $0$B orientifold. The resolution is fairly straightforward, and was already anticipated in \cite{Bergman:1997rf}. 

As argued in \cite{Ginsparg:1986wr}, in toroidal compactifications of the heterotic string the one-loop potential $V(\phi)$ is not flat over moduli space, but is instead extremized at points of enhanced gauge symmetry.  Moreover as argued there at a point $\phi^*$ where the gauge group enhancement has no abelian factors the potential is critical $\nabla V=0$ due to gauge invariance:  $\delta \phi$ is fully charged under the gauge symmetry and in our case they are part of $(\,\vcenter{\hbox{\ydiagram{1,1}}}\,,\vcenter{\hbox{\ydiagram{1,1}}}\,)$.

Expanding $V$ around this vacuum, $\phi = \phi^* + \delta\phi$, the effective potential takes the form
\begin{equation}
V(\phi) = V(\phi^*) + \frac{1}{2}
\left.\frac{\partial^2 V}{\partial \phi_\alpha \partial \phi_\beta}\right|_{\phi^*}
\delta\phi_\alpha \delta\phi_\beta + \cdots \, ,
\end{equation}
so that the fluctuations $\delta\phi_\alpha$ acquire masses
\begin{equation}
m^2_{\alpha\beta} =
\left.\frac{\partial^2 V}{\partial \phi_\alpha \partial \phi_\beta}\right|_{\phi^*}.
\end{equation}

It is natural to expect $V$ is generated as there is no supersymmetry.  All we need to assume is that the eigenvalues of the mass squared matrix are all positive (which we argue heuristically below) leading to their disappearance from the light modes in the strong coupling dual which is proposed to be the orientifold of 0B, rendering the moduli massive and invisible at weak coupling 0B. Indeed, the one-loop cosmological constant in bosonic string theory was computed analytically using the methods in \cite{Baccianti:2025gll}, and is non-vanishing, providing precisely the ingredient needed to lift the moduli.  One finds a correction to $V$ at one loop (in string frame):\footnote{We thank L. Eberhardt for performing this computation based on our request.}
\begin{align}
V= - \int_F \frac{d^2 \tau}{({\rm Im} \tau)^6} \frac{1}{2|\eta^{24}|^2}\left(|\vartheta_2^{16}|^2+|\vartheta_3^{16}|^2+|\vartheta_4^{16}|^2\right) \approx \left(54715 \pm \frac{1408\pi^6i}{5}\right).
\end{align}
Note that the imaginary part denotes the perturbative instability of the tachyon of the bosonic string.  

The Narain moduli are the (Cartan,Cartan) part of the $(Adj,Adj)$ scalar fields of $SO(32)_L\times SO(32)_R$ at the symmetric point.  Let $\phi^{aa'}$ denote the $(Adj,Adj)$ scalar field where $a,a'$ label adjoint elements of $SO(32)_L\times SO(32)_R$.  To compute the mass term we need to compute the quadratic term in the potential.  Gauge invariance uniquely fixes it to be
\begin{align}
    V=\frac{1}{2}A \sum_{a,a'}\phi^{aa'}\phi^{aa'}+\dots
\end{align}
which is completely captured by one parameter $A$.  So at least it is not too unreasonable to assume that this one parameter ends up having a positive value at least for large enough coupling.  At tree level $A=0$, however it will receive corrections as we increase the coupling.
It can be shown that at 1-loop (doing a boost along one direction) $A$ is proportional to
\begin{align}
\begin{split}
    A\propto - \int_F \frac{d^2 \tau}
{({\rm Im} \tau)^6} \frac{1}{2|\eta^{24}|^2}
\left(|\vartheta_2^{15}|^2{\cal D}\cdot |\vartheta_2|^2+|\vartheta_3^{15}|^2{\cal D}\cdot |\vartheta_3|^2+|\vartheta_4^{15}|^2{\cal D}\cdot |\vartheta_4|^2\right)  
\approx - 10^6\ 
\end{split}
\end{align}
where 

\begin{equation}
    {\cal{D}}={\rm log} (q{\overline q})\left(q\frac{\partial}{\partial q}+{\overline q}\frac{\partial}{\partial \overline{q}}\right)+2{\rm log}(q\overline q)^2
\left(q\frac{\partial}{\partial q}{\overline q}\frac{\partial}{\partial \overline{q}}\right).
\end{equation}
Since we find $A<0$ at 1-loop, this shows that at least perturbatively the Narain moduli become tachyonic and in this direction $V''/V\sim -O(1)$.   For the duality to hold up at strong coupling corrections to $A$ should reverse its sign to mass up all Narain moduli which shows that going to strong coupling is a crucial ingredient for this duality to make sense.

\paragraph{Higgsing the B-field and tachyon dynamics}As outlined in the previous section, the Type $0$B orientifold contains two RR two-form fields, $B^\pm$. One linear combination, $B^+$, couples electrically to the $D1^+$-brane, which is identified with the fundamental string in the dual bosonic description. By contrast, the second two-form, $B^-$, does not appear among the massless fields in the dual frame. This raises the question of its fate, as well as that of the $D1^-$-brane to which it couples electrically. In particular, we must understand the mechanism by which both $B^-$ and the corresponding $D1^-$ are removed from the low-energy spectrum.

Let us start by analyzing the tensions of the two D-strings. It was shown in \cite{Klebanov:1998yya} that the tension of stable Type 0 D-branes is given by that of Type II branes divided by $\sqrt{2}$. Furthermore, it was argued in \cite{Klebanov:1998yya,Garousi:1999fu} that $\pm$-type branes couple differently to the closed string tachyon $T_0$. Giving a vev to the tachyon therefore shifts the tensions of the branes. It was conjectured in \cite{Garousi:1999fu} (based on checking its validity including the $O(T^2)$ term) that the string frame D$p$-brane tensions are given by: 
\begin{equation}
    T^{(0)}_{Dp^\mp}= \frac{1}{\sqrt{2}(2 \pi)^p}\frac{M_0^{p+1}}{\lambda_0\sqrt{{1 \pm T_0/2}}}\,,
\end{equation}
where $\lambda_0$ is the string coupling of Type 0B and $M_0$ is its string mass. Assuming that the orientifold does not change the coupling of the D-branes to the closed string tachyon $T_0$, we write the tensions of the $D1^\pm$ and $D9^\pm$ in the 0B orientifold as: 
\begin{equation}
    T_{D1^\mp} \simeq \frac{M_0^{2}}{\lambda_0\sqrt{{1 \pm T_0/2}}}\,,\quad  T_{D9^\mp }\simeq \frac{M_0^{10}}{\lambda_0 \sqrt{{1 \pm T_0/2}}}\,.
\end{equation}
Since both strings couple identically to the dilaton at $T=0$, assuming that the same formulas are valid after orientifolding (which for simplicity we will assume), there is no obvious way to eliminate only one of them simply by making $\lambda_0$ large. This strongly suggests that the tachyon must play a role in the duality map. In particular, the map cannot be as simple as $\lambda_B \sim (\lambda_0)^{-1}$, but must involve a non-trivial transformation of the tachyon. Furthermore, although the background $D9^-$ and anti-$D9^-$ branes allow for the cancellation of the massless dilaton tadpole, the tachyon tadpole, unlike the bosonic side, remains uncancelled. So the duality demands that we move away from $T_0=0$.

Now consider the duality between Type I string theory and the heterotic $SO(32)$ theory. Under this duality, the Type I D1-brane is mapped to the heterotic fundamental string. By analogy, the $D1^{+}$ brane plays the role of the dual bosonic fundamental string in the present setup. Therefore, we expect the ratio $T_{D1^+}/M_{pl}^2$ to become light as the duality is implemented. This behavior can be captured by introducing an effective coupling
\begin{equation}
    \lambda_{eff} = \lambda_0 \sqrt{1 - T_0/2} \, .
\end{equation}
We will be interested in the limit where $T_0\rightarrow 2, \lambda_0\rightarrow \infty$ while keeping $\lambda_{eff}$ fixed.
With this, it is tempting to write a duality map which relates the tension of the fundamental bosonic string to the tension of the $D1^{-}$:
\begin{equation}\label{eq:matchstrings}
    \frac{M_0^2}{\lambda_{eff}}\leftrightarrow  M_B^2\,.
\end{equation}One can also relate the tensions of the five-branes one each side: 
\begin{equation}\label{eq:matchfive}
    \frac{M_0^6}{\lambda_{eff}}\leftrightarrow  \frac{M_B^6}{\lambda_B^2} f(T_B)\,,
\end{equation}
where $f(T_B)$ is an unknown function that describes how the tension of the NS5 brane can be shifted by the bosonic closed string tachyon $T_B$. If we assume that the tachyon sits at its maximum $T_B=0$, then we can approximate $f(T_B)\sim1$  and use \eqref{eq:matchstrings} and \eqref{eq:matchfive} to write: 
\begin{equation}
    \lambda_{eff} \leftrightarrow \lambda _B^{-1}\,.
\end{equation}
While deriving this relation requires a number of assumptions (most notably that the conjecture of \cite{Garousi:1999fu} applies to the Type~0B orientifold and that one can consistently consider the locus where \(f(T_B)=1\)) the resulting relation between the couplings is strikingly similar to that of the familiar Type~I/\(het.\ SO(32)\) duality, suggesting a remarkably coherent picture.

With this in hand we now turn to the $D1^-$ string and the corresponding B-field. Let us consider the dimensionless quantity: 
\begin{equation}
    \frac{T_{D1^-}}{T_{D1^
    +}}= \frac{T_{D1^-}}{M_B^2}=\frac{\sqrt{{1 -T_0/2}}}{\sqrt{{1 + T_0/2}}}\,,
\end{equation}
where in the first equation we have used the fact that the $D1^{+}$ is the bosonic string. We see immediately that the $D1^{-}$ becomes tensionless at $T_0 = 2$. What is the significance of this? We now argue that this is signaling approaching the point which leads to Higgsing of the $B^{-}$ field by $D1^-$ condensate.

Indeed, the appearance of a tensionless string charged under a two-form gauge field $B^{-}$ indicates that the effective description in terms of a weakly coupled abelian tensor field is breaking down. In the simplest scenario, the tensionless limit can lead to condensation of $D1^-$ strings, leading to a higher-form analogue of the Higgs mechanism in which the charged strings condense. In this case, the two-form gauge field $B^{-}$ acquires a mass and the corresponding two-form gauge symmetry is spontaneously broken.

Indeed this type of phenomenon is already well known in supersymmetric backgrounds of string theory as happens for instance in six-dimensional SCFTs with ${\cal N}=(1,0)$ supersymmetry. In such situations, the tensor-field description is no longer sufficient at the transition point, and the correct infrared physics is governed by an interacting conformal fixed point.

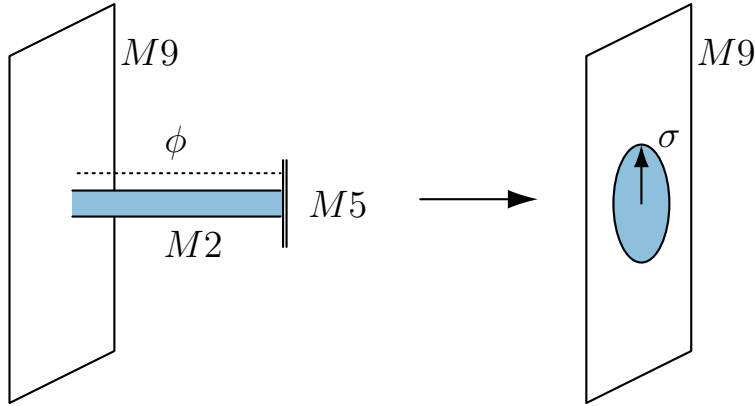
\begin{figure}
    \centering
    
\begin{tikzpicture}[scale=1.15, line cap=round, line join=round]

\begin{scope}

  \filldraw[fill=white!20, draw=black, thick]
    (-0.2,-0.1) -- (1,0.5) -- (1,4.5) -- (-0.2,3.9) -- cycle;
  \node at (1.4,3.95) {\Large $M9$};

  \draw[dotted, thick] (0.58,2.55) -- (2.90,2.55);
  \node at (1.7,2.9) {\Large $\phi$};

\draw[thick] (2.93,1.7) -- (2.93,2.7);
\draw[thick] (2.97,1.7) -- (2.97,2.7);

\node[right] at (3.1,2.2) {\Large $M5$};

  \fill[dark-blue2!45]
    (0.52,2.05) -- (2.92,2.05) -- (2.92,2.35) -- (0.52,2.35) -- cycle;

  \draw[thick] (0.52,2.05) -- (2.90,2.05);
  \draw[thick] (0.52,2.35) -- (2.90,2.35);

  \node[below] at (1.9,2.0) {\Large $M2$};

\end{scope}

\draw[-{Latex[length=4mm,width=2.5mm]}, thick] (4.5,2.25) -- (5.85,2.25);

\begin{scope}[xshift=6.6cm]

  \filldraw[fill=white!20, draw=black, thick]
    (-0.2,-0.1) -- (1,0.5) -- (1,4.5) -- (-0.2,3.9) -- cycle;
  \node at (1.4,3.95) {\Large $M9$};

  \begin{scope}
    \clip (0,0) -- (0.9,0.45) -- (0.9,4.45) -- (0,4) -- cycle;
    \fill[dark-blue2!45] (0.43,2.2) ellipse [x radius=0.32, y radius=0.68];
    \draw[thick]    (0.43,2.2) ellipse [x radius=0.32, y radius=0.68];
  \end{scope}

  \draw[-{Latex[length=3mm,width=2mm]}, thick]
    (0.43,2.2) -- (0.43,2.88);
  \node[right] at (0.50,2.95) {\Large $\sigma$};

\end{scope}

\end{tikzpicture}
    \caption{Higgsing of a 2-form gauge field in M-theory. As the distance $\phi$ between the M5 and M9 shrinks, the M2 stretched in between becomes tensionless. On the M9, it is seen as an instanton, which can be given a size $\sigma$.}
    \label{fig:E8small}
\end{figure}

Let us review some concrete realization of this phenomenon in superstrings/M-theory, like the small instanton transition in heterotic M-theory \cite{Seiberg:1996vs,Ganor:1996mu,Ovrut:2000qi}. In that setup, we consider a CFT related to bringing an M5 brane to the Ho\v{r}ava-Witten M9 boundary. There is an M2 brane stretching between an M5 brane in the bulk and the boundary M9. The distance between the M5 and the M9 is parametrized by a modulus $\varphi$ (the scalar field of a tensor multiplet living on the M5 brane). This modulus also parametrizes the tension of the E-string on the M5 brane worldvolume, since it comes from the M2 stretching between the M5 and the boundary. As one brings the M5 close to the M9, the E-string becomes tensionless at $\varphi=0$. At this point a new branch opens up: the Higgs branch, where the 2-form $B$ on the M5 gets massed up and the tensor multiplet is swapped for hypermultiplets which describe the moduli of an $E_8$ instanton. An M5 brane at the wall corresponds to 0 size instanton.  Giving a vev to hypers which controls the size of the instanton puffs it into a finite-sized $E_8$ instanton. The size of this $E_8$ instanton is parametrized by a modulus $\sigma$ (among other instanton moduli), as $B$ is massed up along the Higgs branch. As instanton gets puffed up, M5 brane is smeared on the M9 brane and this leads to Higgsing of the B-field which was living on it and also there is no more the corresponding tensionless string, see \autoref{fig:E8small}.
This proves that there is a natural mechanisms by which a B-field can be Higgsed by the presence of a tensionless string that is electrically charged under it. 

\begin{figure}
    \centering
    \includegraphics[width=0.8\linewidth]{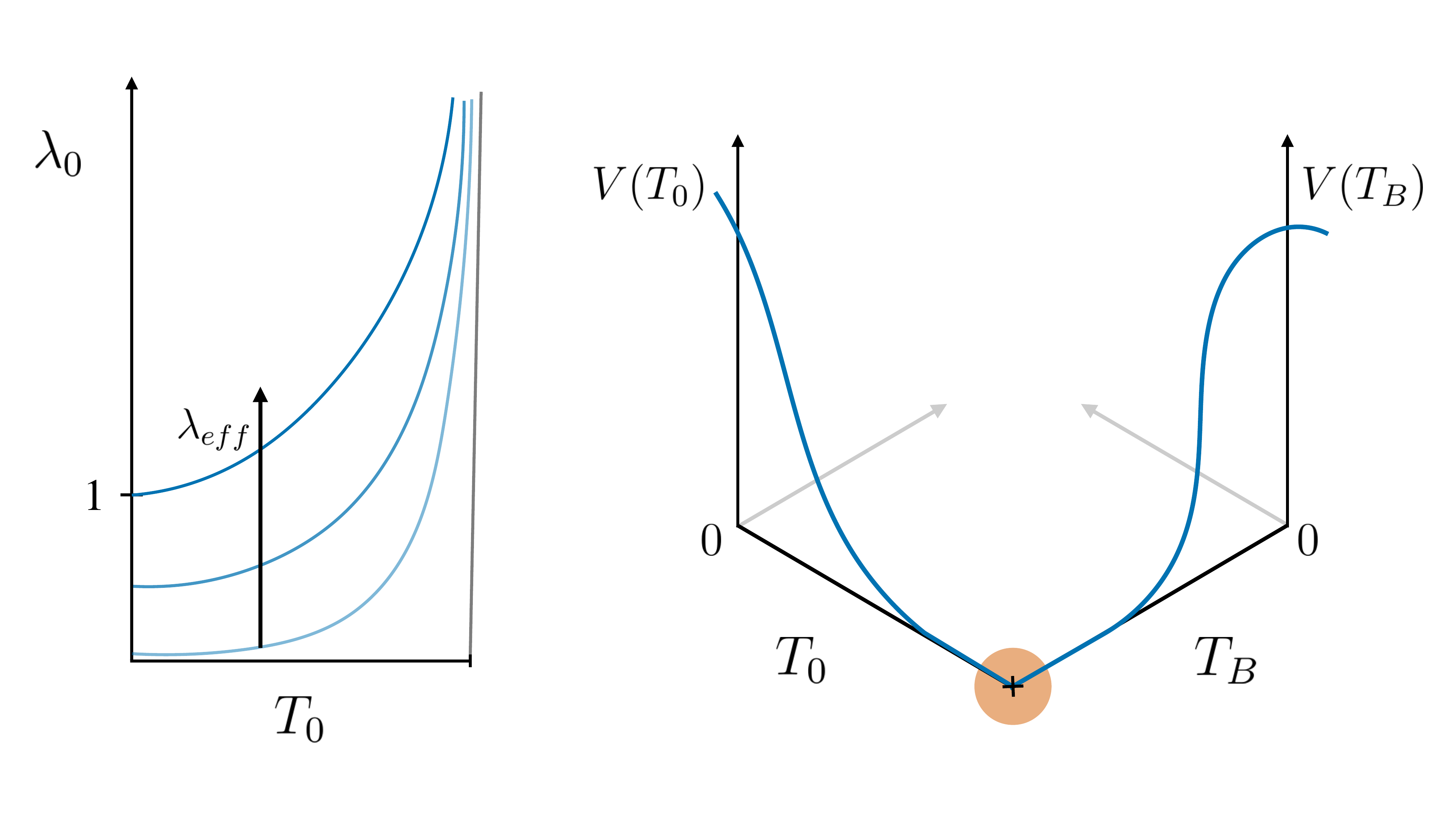}
    \caption{On the left, we draw $\lambda_{eff}=const$ contours in the $\{\lambda_0, T_0\}$ field space. On the right, we illustrate how condensing the closed string tachyon on either sides of the duality could lead to a strongly coupled interface between the two theories, where additional degrees of freedom become massless.  On the 0B side tuning $T_0$ gets us to the tensionless $D1^-$ string, on the bosonic side giving vev to $T_B$ should correspond to un-Higgsing of a massive rank 2 tensor field leading to a massless $B^-$ at the transition point. }
    \label{fig:tachyons}
\end{figure}

It could therefore be the case for us that the $D1^-$ becoming tensionless is signaling the approach to the point where we can accomplish the Higgsing of the extra $B^-$-field on the 0B side of the duality, where the scalar $1-T_0/2$ plays the role of $\varphi$.  Thus $T_0=2$ plays the analogous role to $\varphi =0$ in the small instanton story.  To move away and Higgs the anti-symmetric tensor field one would also need a field like $\sigma$.  Here we also need such a field that accomplishes the Higgsing of the $B^-$ field after we have reached the $T_0=2$ point, corresponding to condensing the $D1^-$ branes.  It is natural to believe that the bosonic string tachyon $T_B$ plays the role of $\sigma$. Namely the transition point should be at a $T_B=a\not =0$ (it is possible that we also may have to adjust the bosonic string coupling to large values to get to this point, but we do not have an apriori reason for that). Changing the value of $T_B$ from $a$ back to $0$ would bring us to the perturbative bosonic string without tachyon condensation.  This is depicted in \autoref{fig:tachyons}.

There is, however, an important distinction from the supersymmetric six-dimensional SCFT case. There, both scalar fields are massless moduli, and one can move smoothly between the tensor and Higgs branches. In the present non-supersymmetric setting, by contrast, neither direction appears to correspond to a flat direction. It is natural to expect that where they meet the potential for the scalar fields $T_0,T_B$ becomes flat as we may expect a non-trivial CFT at that point due to appearance of tensionless strings, though it is not clear if it is possible to decouple it from gravity.

\subsubsection{Further Evidence}\label{sec:BGfurther}
In the previous sections, we reviewed the duality proposed in \cite{Bergman:1997rf}, highlighting several apparent puzzles and proposing their resolutions. We now turn to new evidence in support of this duality.

\paragraph{Charged Tachyon condensation} Up until now we have only discussed the condensation of the neutral closed string tachyons. We now turn to the open string tachyons in the $(\mathbf{32},\mathbf{32})$. We argue that their condensation leads to similar physics on both sides of the duality. On the orientifold side, the $(\mathbf{32},\mathbf{32})$ tachyon comes from open strings stretching between the 9-branes and anti-9-branes. One expects that condensing it would lead to the annihilation of the branes against the anti-branes, leaving behind the orientifold background with massless tadpoles for $\phi$ and the singlet tachyon $T_0$. Indeed, recall from \autoref{sec:0BOmega} that this orientifold does not carry RR charge, so this background is consistent from the perspective of RR tadpole cancellation. Notably, the vacuum energy is negative after the annihilation of the 9-branes, due to the negative contribution from the orientifold to the vacuum energy. 

Let us now turn to the bosonic side and ask what happens upon condensing the $(\mathbf{32},\mathbf{32})$ tachyon. Recall that the 16 compact left- and right-moving bosons on the worldsheet, corresponding to the torus directions, can be equivalently described in terms of 32 real left-moving fermions $\lambda^A(z)$ and 32 real right-moving fermions $\tilde{\lambda}^{\tilde A}(\bar z)$, which is indeed what the $D1^+$ string sees. The $(\mathbf{32},\mathbf{32})$ tachyon arises from the vector primaries of the two affine algebras. At zero momentum, its vertex operators take the form
\begin{equation}
    V_{A \tilde{A}}(z,\bar z) \sim \lambda^A(z)\,\tilde{\lambda}^{\tilde A}(\bar z)\,.
\end{equation}
Turning on a vacuum expectation value for this tachyon corresponds to perturbing the worldsheet action by
\begin{equation}\label{eq:condensate}
    \delta S = \int d^2 z \; T_{A\tilde A}\,\lambda^A(z)\,\tilde{\lambda}^{\tilde A}(\bar z)\,,
\end{equation}
where $T_{A\tilde A}$ transforms in the $(\mathbf{32},\mathbf{32})$. This deformation is naturally interpreted as a mass matrix for the worldsheet fermions. In other words, generic charged tachyon condensate directly generates masses for the 32 left- and 32 right-moving fermions.\footnote{In fact a further evidence is that on both sides we can go in the direction of gradually reducing the gauge theory ranks from $SO(32)\times SO(32)\rightarrow SO(32-k)\times SO(32-k)$, where on the 0B side we annihilate $k$ of the brane/anti-brane pairs, and from the bosonic side it corresponds to the worldsheet description where we give mass to $k$ pairs of left/right fermions.} As a result, these degrees of freedom decouple at low energies, effectively removing the corresponding 16 compact spacetime directions associated with the torus. From the worldsheet point of view, this has an immediate consequence for the central charge. Before condensation, the 32 left- and 32 right-moving fermions contribute
\begin{equation}
c_L = c_R = 16 \, ,
\end{equation}
which precisely accounts for the central charge of the 16 compact bosons. Once these fermions acquire a mass, they no longer contribute to the infrared conformal field theory, leading to a deficit in the total central charge. In the spacetime effective action this manifests itself as a non-vanishing dilaton gradient
and an associated negative contribution to the vacuum energy as is the case in the 0B side:
\begin{equation}
S_{\rm bosonic}
\sim
\frac{1}{2\kappa^2}\int d^{10} x\,\sqrt{-g}\,e^{-2\Phi}
\left[
R + 4(\partial\Phi)^2 + \frac{2\delta c}{3\alpha'} + \cdots
\right] \,,
\end{equation}
where $\delta c = 26-D = 16$.

\paragraph{Other charged branes and K-theory} So far, we have focused on the $D1^\pm$ branes in the orientifold of Type $0\mathrm{B}$, which carry integral RR charge. The theory also contains magnetically charged $D5^\pm$ branes, and it is natural to ask how these map under the proposed duality. By analogy with the fate of the $D1^\pm$ branes, one is led to expect that the $D5^{-}$ maps to the NS21-brane wrapped on the $T^{16}$ on the bosonic side, while the $D5^{+}$ becomes tensionless in bosonic string units. However, this is not the full story. As in Type I string theory, there can also exist branes carrying torsion-valued RR charges, which are classified by K-theory. The relevant K-theory groups for this setup were determined in \cite{Kaidi:2019tyf}.

In close analogy with the Type I case, one finds additional non-BPS branes carrying $\mathbb{Z}_2$ charges. In particular, there are two $\widehat{D(-1)}$, two $\widehat{D0}$, two $\widehat{D7}$, and two $\widehat{D8}$ branes, where the doubling reflects the two RR sectors of the Type $0$ theory. Each of these carries a $\mathbb{Z}_2$ K-theory charge. As in the analogous discussion of Type I/heterotic duality in \cite{Witten:1998cd}, a natural question is whether these states admit counterparts on the bosonic side. This issue was addressed in \cite{Michishita:1999it}, where it was proposed that the $\widehat{D0}$ branes correspond to massive states transforming in both the $(s,s)$ and $(c,c)$ representations of $Spin(32)\times Spin(32)$ in the bosonic theory (thus making the gauge group $(Spin(32)\times Spin(32))/\mathbb{Z}_2^2$) providing further evidence for the BG conjecture. A more detailed investigation of these questions would be worthwhile, and we defer it to future work.

\subsection{DMS Duality}
\label{sec:DMS}

Mourad, Sagnotti and one of us (DMS) have proposed in \cite{Dudas:2001wd} that the orientifold of the Type 0A string with vanishing dilaton and open string singlet tachyon tadpoles is S-dual to the orientifold of the bosonic string on the symmetric $SO(32)_L\otimes SO(32)_R$ lattice (where the orientifold exchanges the two lattices). The evidence for this proposal is similar to that for the BG duality discussed in the previous subsection.  The gauge group is $SO(32)$ on both sides and the matter content on both sides is given in \autoref{fig:0A-bosonic}.  The $D1^+$brane in this setup is critical, and its field content matches that of the
corresponding non-oriented bosonic string. 

\begin{table}[h!]
    \centering
    \[
\begin{array}{c|c}
0A/\Omega & \text{Bosonic String on } T^{16}_{SO(32)_L\ \otimes SO(32)_R}/\Omega \\
\hline
g_{\mu\nu},\phi & g_{\mu\nu},\phi \\
A_{\mu} & A_{\mu}\\
\mathbf{1}~T & \mathbf{1}~T \\
\ydiagram{2} ~T & \ydiagram{2}~T \\
A^+,C^+ & \\
& (\vcenter{\hbox{~\ydiagram{1,1}}}\otimes \vcenter{\hbox{~\ydiagram{1,1}}}\,)_{Sym}\supset~\text{Narain moduli}
\end{array}
\]
    \caption{Massless and tachyonic spectrum of $0A/\Omega$. There is a mismatch by an extra 1-form and a 3-form field on the $0A/\Omega$ side and (tree-level) massless scalars in the Bosonic orientifold side.}
    \label{fig:0A-bosonic}
\end{table}

As can be seen from the table there are two mismatches, the fields which leads to Narain moduli as well as $A^+,C^+$.

The argument for Narain moduli picking up mass is the same as the BG case where the potential is generated which masses them up.  

The 1-form $A^+$ and 3-form $C^+$ mismatch  goes away by the tachyon condensation which shrinks the $S^{1+}$ circle and removes the $A^+,C^+$ as was discussed in \cite{Baykara:2026gem}.  The tachyon field does not survive this shrinking.  Similarly we can expect that the singlet tachyon of the bosonic side picks up mass at some vev for $T_B$ and is removed from the light modes similar to the Narain moduli.\footnote{Note that on the bosonic side, the orientifold gives rise to 
a non-zero tachyonic tadpole for the singlet in the symmetric representation of $SO(32)$, but this is not required to match the energy on the other side, since as we vary the coupling $V$ changes.}
We would naively be led to identifying similar to the $0B/\Omega$ and bosonic string the couplings as 
\begin{align}
    \lambda_0\sim 1/\lambda_B.
\end{align}
At the end we would be left, after tachyon condensation for 0A orientifold, with M-theory on a $\bZ_2$ quotient of $S^1$ with 32 M10 branes wrapping it and the duality would imply that this is dual to $SO(32)_L\otimes SO(32)_R$ bosonic string orientifold with a specific vev for the bosonic string tachyon giving it a mass.  This suggests that perhaps the relation between $\lambda_0,\lambda_B$ is more complicated and it would involve the choice of the bosonic tachyon field as well.  This would be needed if we wish to avoid identifying large circle M-theory orbifold with M10 branes wrapping it with a weak coupling bosonic string (as we would not expect to have both large radius M-theory and weak coupling bosonic descriptions at the same time).

\section{Concluding Thoughts}\label{sec:conclude}

In this paper we have found an intricate web of dualities between non-supersymmetric strings.  This includes new ones as well as finding evidence and resolving puzzles of some of the previously proposed dualities.  Using the new ones we have gained further insight into the strong coupling description of some of them and in particular that of $SO(16)\times SO(16)$ non-supersymmetric tachyon-free theory.

We have not considered all possible quotients of $S^1\vee S^1$ or $(S^1\vee S^1) \times S^1$ and focused mostly on the ones we can identify with those that have some weak coupling string theory descriptions.  One natural thing is to extend this list and include addition quotients including $\bZ_2\times \bZ_2$ quotients.

We have also not considered the versions of the orientifolds where the gauge symmetries are projected to symplectic rather than orthogonal groups. This includes Sugimoto string with gauge group $USp(32)$ \cite{Sugimoto:1999tx}, as well as the 0A and 0B orientifolds except Sagnotti's. Sugimoto string is based on Type IIB and so has no connection with Type 0 strings. It was conjectured in \cite{Angelantonj:2007ts} to be nonperturbatively interpreted as a metastable vacuum of Type I string.  In general, the difference between these orientifolds is expected to be non-geometric and related to choice of signs in the tadpole diagrams.

Clearly a lot remains to be done, but we have provided what we feel is strong evidence that duality symmetries in quantum gravitational theories do not require supersymmetry and can lead to insights also in the more interesting and physically relevant case of non-supersymmetric quantum gravitational theories, as is the case for our universe.

\acknowledgments
We would like to thank G. Bossard, L. Eberhardt, Y. Hamada, M. Montero, I. Valenzuela for valuable discussions.

The work of ZKB, MD, HPDF and CV is supported in part by a grant from the Simons Foundation (602883,CV) and a gift from the DellaPietra Foundation. The work of E.~D. was supported in part by the IRP UCMN France-USA.

\bibliographystyle{JHEP}
\bibliography{biblio.bib}

@article{Bergman:1999km,
    author = "Bergman, Oren and Gaberdiel, Matthias R.",
    title = "{Dualities of type 0 strings}",
    eprint = "hep-th/9906055",
    archivePrefix = "arXiv",
    reportNumber = "CALT-68-2228, DAMTP-1999-74",
    doi = "10.1088/1126-6708/1999/07/022",
    journal = "JHEP",
    volume = "07",
    pages = "022",
    year = "1999"
}

@article{Polchinski:1995df,
    author = "Polchinski, Joseph and Witten, Edward",
    title = "{Evidence for heterotic - type I string duality}",
    eprint = "hep-th/9510169",
    archivePrefix = "arXiv",
    reportNumber = "IASSNS-HEP-95-81, NSF-ITP-95-135",
    doi = "10.1016/0550-3213(95)00614-1",
    journal = "Nucl. Phys. B",
    volume = "460",
    pages = "525--540",
    year = "1996"
}

@article{Kaidi:2019tyf,
    author = "Kaidi, Justin and Parra-Martinez, Julio and Tachikawa, Yuji",
    title = "{Topological Superconductors on Superstring Worldsheets}",
    eprint = "1911.11780",
    archivePrefix = "arXiv",
    primaryClass = "hep-th",
    reportNumber = "IPMU-19-0164, UCLA/TEP/2019/106",
    doi = "10.21468/SciPostPhys.9.1.010",
    journal = "SciPost Phys.",
    volume = "9",
    pages = "10",
    year = "2020"
}

@article{Angius:2022mgh,
    author = "Angius, Roberta and Delgado, Matilda and Uranga, Angel M.",
    title = "{Dynamical Cobordism and the beginning of time: supercritical strings and tachyon condensation}",
    eprint = "2207.13108",
    archivePrefix = "arXiv",
    primaryClass = "hep-th",
    doi = "10.1007/JHEP08(2022)285",
    journal = "JHEP",
    volume = "08",
    pages = "285",
    year = "2022"
}

@article{Basile:2023knk,
    author = "Basile, Ivano and Debray, Arun and Delgado, Matilda and Montero, Miguel",
    title = "{Global anomalies {\&} bordism of non-supersymmetric strings}",
    eprint = "2310.06895",
    archivePrefix = "arXiv",
    primaryClass = "hep-th",
    reportNumber = "IFT-23-129",
    doi = "10.1007/JHEP02(2024)092",
    journal = "JHEP",
    volume = "02",
    pages = "092",
    year = "2024"
}

@article{Fabinger:2000jd,
    author = "Fabinger, Michal and Horava, Petr",
    title = "{Casimir effect between world branes in heterotic M theory}",
    eprint = "hep-th/0002073",
    archivePrefix = "arXiv",
    reportNumber = "CALT-68-2255, CITUSC-00-004, PRA-HEP-00-02",
    doi = "10.1016/S0550-3213(00)00255-8",
    journal = "Nucl. Phys. B",
    volume = "580",
    pages = "243--263",
    year = "2000"
}

@article{Hellerman:2004zm,
    author = "Hellerman, Simeon",
    title = "{On the landscape of superstring theory in D {\ensuremath{>}} 10}",
    eprint = "hep-th/0405041",
    archivePrefix = "arXiv",
    month = "5",
    year = "2004"
}

@article{Anastasi:2026cus,
    author = "Anastasi, Edoardo and Montero, Miguel and Uranga, Angel M. and Wang, Chuying",
    title = "{What IIB looks IIA string: String Cobordisms via Non-Compact CFTs}",
    eprint = "2603.00225",
    archivePrefix = "arXiv",
    primaryClass = "hep-th",
    reportNumber = "IFT-26-016",
    month = "2",
    year = "2026"
}

@article{Altavista:2026edv,
    author = "Altavista, Chiara and Anastasi, Edoardo and Angius, Roberta and Uranga, Angel M.",
    title = "{The Art of Branching: Cobordism Junctions of 10d String Theories}",
    eprint = "2603.24667",
    archivePrefix = "arXiv",
    primaryClass = "hep-th",
    month = "3",
    year = "2026"
}

@article{Kaidi:2024cbx,
    author = "Kaidi, Justin and Tachikawa, Yuji and Yonekura, Kazuya",
    title = "{On non-supersymmetric heterotic branes}",
    eprint = "2411.04344",
    archivePrefix = "arXiv",
    primaryClass = "hep-th",
    reportNumber = "TU-1248, KYUSHU-HET-294",
    doi = "10.1007/JHEP03(2025)211",
    journal = "JHEP",
    volume = "03",
    pages = "211",
    year = "2025"
}

@article{Hellerman:2006ff,
    author = "Hellerman, Simeon and Swanson, Ian",
    title = "{Dimension-changing exact solutions of string theory}",
    eprint = "hep-th/0612051",
    archivePrefix = "arXiv",
    doi = "10.1088/1126-6708/2007/09/096",
    journal = "JHEP",
    volume = "09",
    pages = "096",
    year = "2007"
}

@article{Kaidi:2020jla,
    author = "Kaidi, Justin",
    title = "{Stable Vacua for Tachyonic Strings}",
    eprint = "2010.10521",
    archivePrefix = "arXiv",
    primaryClass = "hep-th",
    doi = "10.1103/PhysRevD.103.106026",
    journal = "Phys. Rev. D",
    volume = "103",
    number = "10",
    pages = "106026",
    year = "2021"
}

@article{Hellerman:2006nx,
    author = "Hellerman, Simeon and Swanson, Ian",
    title = "{Cosmological solutions of supercritical string theory}",
    eprint = "hep-th/0611317",
    archivePrefix = "arXiv",
    doi = "10.1103/PhysRevD.77.126011",
    journal = "Phys. Rev. D",
    volume = "77",
    pages = "126011",
    year = "2008"
}

@article{Hellerman:2007fc,
    author = "Hellerman, Simeon and Swanson, Ian",
    title = "{Charting the landscape of supercritical string theory}",
    eprint = "0705.0980",
    archivePrefix = "arXiv",
    primaryClass = "hep-th",
    doi = "10.1103/PhysRevLett.99.171601",
    journal = "Phys. Rev. Lett.",
    volume = "99",
    pages = "171601",
    year = "2007"
}

@article{Altavista:2026evd,
    author = "Altavista, Chiara and Anastasi, Edoardo and Raucci, Salvatore and Uranga, Angel M. and Wang, Chuying",
    title = "{Ho{\v{r}}ava-Witten theory on ${\mathbf{S}}^1\vee{\mathbf{S}}^1$ as type 0 orientifold}",
    eprint = "2603.25786",
    archivePrefix = "arXiv",
    primaryClass = "hep-th",
    reportNumber = "IFT-UAM/CSIC-26-39",
    month = "3",
    year = "2026"
}

@article{Alvarez-Gaume:1984zlq,
    author = "Alvarez-Gaume, Luis and Ginsparg, Paul H.",
    editor = "Salam, A. and Sezgin, E.",
    title = "{The Structure of Gauge and Gravitational Anomalies}",
    reportNumber = "HUTP-84/A016",
    doi = "10.1016/0003-4916(85)90087-9",
    journal = "Annals Phys.",
    volume = "161",
    pages = "423",
    year = "1985",
    note = "[Erratum: Annals Phys. 171, 233 (1986)]"
}

@article{Witten:1998cd,
    author = "Witten, Edward",
    title = "{D-branes and K-theory}",
    eprint = "hep-th/9810188",
    archivePrefix = "arXiv",
    reportNumber = "IASSNS-HEP-98-82",
    doi = "10.1088/1126-6708/1998/12/019",
    journal = "JHEP",
    volume = "12",
    pages = "019",
    year = "1998"
}

@article{Michishita:1999it,
    author = "Michishita, Yoji",
    title = "{D0-branes in SO(32) x SO(32) open type 0 string theory}",
    eprint = "hep-th/9907094",
    archivePrefix = "arXiv",
    reportNumber = "KUNS-1586",
    doi = "10.1016/S0370-2693(99)01123-5",
    journal = "Phys. Lett. B",
    volume = "466",
    pages = "161",
    year = "1999"
}

@article{Klebanov:1998yya,
    author = "Klebanov, Igor R. and Tseytlin, Arkady A.",
    title = "{D-branes and dual gauge theories in type 0 strings}",
    eprint = "hep-th/9811035",
    archivePrefix = "arXiv",
    reportNumber = "PUPT-1819, IMPERIAL-TP-98-99-07",
    doi = "10.1016/S0550-3213(99)00041-3",
    journal = "Nucl. Phys. B",
    volume = "546",
    pages = "155--181",
    year = "1999"
}

@article{Baccianti:2025gll,
    author = "Baccianti, Marco Maria and Chandra, Jeevan and Eberhardt, Lorenz and Hartman, Thomas and Mizera, Sebastian",
    title = "{Rademacher expansion of modular integrals}",
    eprint = "2501.13827",
    archivePrefix = "arXiv",
    primaryClass = "hep-th",
    doi = "10.21468/SciPostPhys.19.4.103",
    journal = "SciPost Phys.",
    volume = "19",
    number = "4",
    pages = "103",
    year = "2025"
}

@article{Dudas:2001wd,
    author = "Dudas, E. and Mourad, J. and Sagnotti, A.",
    title = "{Charged and uncharged D-branes in various string theories}",
    eprint = "hep-th/0107081",
    archivePrefix = "arXiv",
    reportNumber = "LPT-ORSAY-01-56, ROM2F-01-18",
    doi = "10.1016/S0550-3213(01)00552-1",
    journal = "Nucl. Phys. B",
    volume = "620",
    pages = "109--151",
    year = "2002"
}

@article{Seiberg:1996vs,
    author = "Seiberg, N. and Witten, Edward",
    title = "{Comments on string dynamics in six-dimensions}",
    eprint = "hep-th/9603003",
    archivePrefix = "arXiv",
    reportNumber = "RU-96-12, IASSNS-HEP-96-19",
    doi = "10.1016/0550-3213(96)00189-7",
    journal = "Nucl. Phys. B",
    volume = "471",
    pages = "121--134",
    year = "1996"
}

@article{Ganor:1996mu,
    author = "Ganor, Ori J. and Hanany, Amihay",
    title = "{Small E(8) instantons and tensionless noncritical strings}",
    eprint = "hep-th/9602120",
    archivePrefix = "arXiv",
    reportNumber = "IASSNS-HEP-96-12, PUPT-1595",
    doi = "10.1016/0550-3213(96)00243-X",
    journal = "Nucl. Phys. B",
    volume = "474",
    pages = "122--140",
    year = "1996"
}

@article{Ovrut:2000qi,
    author = "Ovrut, Burt A. and Pantev, Tony and Park, Jaemo",
    title = "{Small instanton transitions in heterotic M theory}",
    eprint = "hep-th/0001133",
    archivePrefix = "arXiv",
    reportNumber = "UPR-871-T, OUTP-99-03P, IASSNS-HEP-00-03",
    doi = "10.1088/1126-6708/2000/05/045",
    journal = "JHEP",
    volume = "05",
    pages = "045",
    year = "2000"
}

@article{Bergman:1997rf,
    author = "Bergman, Oren and Gaberdiel, Matthias R.",
    title = "{A Nonsupersymmetric open string theory and S duality}",
    eprint = "hep-th/9701137",
    archivePrefix = "arXiv",
    reportNumber = "HUTP-97-A003, BRX-TH-402",
    doi = "10.1016/S0550-3213(97)00309-X",
    journal = "Nucl. Phys. B",
    volume = "499",
    pages = "183--204",
    year = "1997"
}

@article{Houart:2000vm,
    author = "Houart, Laurent and Lozano, Yolanda",
    title = "{Brane descent relations in M theory}",
    eprint = "hep-th/0001170",
    archivePrefix = "arXiv",
    reportNumber = "CERN-TH-2000-025, IMPERIAL-TP-99-00-19",
    doi = "10.1016/S0370-2693(00)00317-8",
    journal = "Phys. Lett. B",
    volume = "479",
    pages = "299--307",
    year = "2000"
}

@article{Baykara:2026gem,
    author = "Baykara, Zihni Kaan and Dudas, Emilian and Vafa, Cumrun",
    title = "{M-theory on $S^1\vee S^1$ as Type 0A}",
    eprint = "2603.13468",
    archivePrefix = "arXiv",
    primaryClass = "hep-th",
    month = "3",
    year = "2026"
}

@article{Garousi:1999fu,
    author = "Garousi, Mohammad R.",
    title = "{String scattering from D-branes in type 0 theories}",
    eprint = "hep-th/9901085",
    archivePrefix = "arXiv",
    reportNumber = "IPM-P-99-5",
    doi = "10.1016/S0550-3213(99)00152-2",
    journal = "Nucl. Phys. B",
    volume = "550",
    pages = "225--237",
    year = "1999"
}

@article{Ginsparg:1986wr,
    author = "Ginsparg, Paul H. and Vafa, C.",
    title = "{Toroidal Compactification of Nonsupersymmetric Heterotic Strings}",
    reportNumber = "HUTP-86-A064A, HUTP-86-A064",
    doi = "10.1016/0550-3213(87)90387-7",
    journal = "Nucl. Phys. B",
    volume = "289",
    pages = "414",
    year = "1987"
}

@article{hw1,
    author = "Horava, Petr and Witten, Edward",
    title = "{Heterotic and Type I string dynamics from eleven dimensions}",
    eprint = "hep-th/9510209",
    archivePrefix = "arXiv",
    reportNumber = "IASSNS-HEP-95-86, PUPT-1571A",
    doi = "10.1016/0550-3213(95)00621-4",
    journal = "Nucl. Phys. B",
    volume = "460",
    pages = "506--524",
    year = "1996"
}

@article{Fraiman:2025yrx,
    author = "Fraiman, Bernardo and Parra de Freitas, H{\'e}ctor",
    title = "{Symmetries and dualities in non-supersymmetric CHL strings}",
    eprint = "2511.01674",
    archivePrefix = "arXiv",
    primaryClass = "hep-th",
    reportNumber = "MPP-2025-79; IFT-UAM/CSIC-25-111",
    month = "11",
    year = "2025"
}

@article{Blumenhagen:1999ad,
    author = "Blumenhagen, Ralph and Kumar, Alok",
    title = "{A Note on orientifolds and dualities of type 0B string theory}",
    eprint = "hep-th/9906234",
    archivePrefix = "arXiv",
    reportNumber = "HUB-EP-99-29, CERN-TH-99-190",
    doi = "10.1016/S0370-2693(99)01002-3",
    journal = "Phys. Lett. B",
    volume = "464",
    pages = "46--52",
    year = "1999"
}

@article{Kachru:1998yy,
    author = "Kachru, Shamit and Silverstein, Eva",
    title = "{Selfdual nonsupersymmetric type II string compactifications}",
    eprint = "hep-th/9808056",
    archivePrefix = "arXiv",
    reportNumber = "SLAC-PUB-7907, LBL-42139, LBNL-42139, UCB-PTH-98-40",
    doi = "10.1088/1126-6708/1998/11/001",
    journal = "JHEP",
    volume = "11",
    pages = "001",
    year = "1998"
}

@article{Blum:1997cs,
    author = "Blum, Julie D. and Dienes, Keith R.",
    title = "{Duality without supersymmetry: The Case of the SO(16) x SO(16) string}",
    eprint = "hep-th/9707148",
    archivePrefix = "arXiv",
    reportNumber = "IASSNS-HEP-97-67",
    doi = "10.1016/S0370-2693(97)01172-6",
    journal = "Phys. Lett. B",
    volume = "414",
    pages = "260--268",
    year = "1997"
}

@article{hw2,
    author = "Horava, Petr and Witten, Edward",
    title = "{Eleven-dimensional supergravity on a manifold with boundary}",
    eprint = "hep-th/9603142",
    archivePrefix = "arXiv",
    reportNumber = "IASSNS-HEP-96-17, PUPT-1597",
    doi = "10.1016/0550-3213(96)00308-2",
    journal = "Nucl. Phys. B",
    volume = "475",
    pages = "94--114",
    year = "1996"
}

@article{Blum:1997gw,
    author = "Blum, Julie D. and Dienes, Keith R.",
    title = "{Strong / weak coupling duality relations for nonsupersymmetric string theories}",
    eprint = "hep-th/9707160",
    archivePrefix = "arXiv",
    reportNumber = "IASSNS-HEP-97-80",
    doi = "10.1016/S0550-3213(97)00803-1",
    journal = "Nucl. Phys. B",
    volume = "516",
    pages = "83--159",
    year = "1998"
}

@article{Acharya:2022shu,
    author = "Acharya, Bobby Samir and Aldazabal, Gerardo and Font, Anamar{\'\i}a and Narain, Kumar and Zadeh, Ida G.",
    title = "{Heterotic strings on $ T^{3}/\mathbb{Z}_{2}$, Nikulin involutions and M-theory}",
    eprint = "2205.09764",
    archivePrefix = "arXiv",
    primaryClass = "hep-th",
    doi = "10.1007/JHEP09(2022)209",
    journal = "JHEP",
    volume = "09",
    pages = "209",
    year = "2022"
}

@article{Alvarez-Gaume:1986ghj,
    author = "Alvarez-Gaume, Luis and Ginsparg, Paul H. and Moore, Gregory W. and Vafa, C.",
    title = "{An O(16) $\times$ O(16) Heterotic String}",
    reportNumber = "HUTP-86/A013",
    doi = "10.1016/0370-2693(86)91524-8",
    journal = "Phys. Lett. B",
    volume = "171",
    pages = "155--162",
    year = "1986"
}

@article{Dixon:1986iz,
    author = "Dixon, Lance J. and Harvey, Jeffrey A.",
    editor = "Schellekens, B.",
    title = "{String Theories in Ten-Dimensions Without Space-Time Supersymmetry}",
    reportNumber = "PRINT-86-0244 (PRINCETON)",
    doi = "10.1016/0550-3213(86)90619-X",
    journal = "Nucl. Phys. B",
    volume = "274",
    pages = "93--105",
    year = "1986"
}

@article{Faraggi:2007tj,
    author = "Faraggi, Alon E. and Tsulaia, Mirian",
    title = "{On the Low Energy Spectra of the Nonsupersymmetric Heterotic String Theories}",
    eprint = "0706.1649",
    archivePrefix = "arXiv",
    primaryClass = "hep-th",
    reportNumber = "LTH-746",
    doi = "10.1140/epjc/s10052-008-0545-2",
    journal = "Eur. Phys. J. C",
    volume = "54",
    pages = "495--500",
    year = "2008"
}

@article{Montero:2025ayi,
    author = "Montero, Miguel and Zapata, Luis",
    title = "{M-theory boundaries beyond supersymmetry}",
    eprint = "2504.06985",
    archivePrefix = "arXiv",
    primaryClass = "hep-th",
    reportNumber = "IFT-25-033",
    doi = "10.1007/JHEP07(2025)090",
    journal = "JHEP",
    volume = "07",
    pages = "090",
    year = "2025"
}

@inproceedings{Sagnotti:1995ga,
    author = "Sagnotti, Augusto",
    title = "{Some properties of open string theories}",
    booktitle = "{International Workshop on Supersymmetry and Unification of Fundamental Interactions (SUSY 95)}",
    eprint = "hep-th/9509080",
    archivePrefix = "arXiv",
    reportNumber = "ROM2F-95-18",
    pages = "473--484",
    month = "9",
    year = "1995"
}

@article{Sugimoto:1999tx,
    author = "Sugimoto, Shigeki",
    title = "{Anomaly cancellations in type I D-9 - anti-D-9 system and the USp(32) string theory}",
    eprint = "hep-th/9905159",
    archivePrefix = "arXiv",
    reportNumber = "YITP-99-25",
    doi = "10.1143/PTP.102.685",
    journal = "Prog. Theor. Phys.",
    volume = "102",
    pages = "685--699",
    year = "1999"
}

@article{bianchi-sagnotti,
    author = "Bianchi, Massimo and Sagnotti, Augusto",
    title = "{On the systematics of open string theories}",
    reportNumber = "ROM2F-90-20",
    doi = "10.1016/0370-2693(90)91894-H",
    journal = "Phys. Lett. B",
    volume = "247",
    pages = "517--524",
    year = "1990"
}

@inproceedings{augusto,
    author = "Sagnotti, Augusto",
    title = "{Open Strings and their Symmetry Groups}",
    booktitle = "{NATO Advanced Summer Institute on Nonperturbative Quantum Field Theory (Cargese Summer Institute)}",
    eprint = "hep-th/0208020",
    archivePrefix = "arXiv",
    reportNumber = "ROM2F-87-25",
    month = "9",
    year = "1987"
}

@article{bcd,
    author = "Bossard, Guillaume and Casagrande, Gabriele and Dudas, Emilian",
    title = "{Twisted orientifold planes and S-duality without supersymmetry}",
    eprint = "2411.00955",
    archivePrefix = "arXiv",
    primaryClass = "hep-th",
    reportNumber = "CPHT-RR012.072024",
    doi = "10.1007/JHEP02(2025)062",
    journal = "JHEP",
    volume = "02",
    pages = "062",
    year = "2025"
}

@article{nonsusystringsrevs1,
    author = "Angelantonj, Carlo and Sagnotti, Augusto",
    title = "{Open strings}",
    eprint = "hep-th/0204089",
    archivePrefix = "arXiv",
    reportNumber = "CERN-TH-2002-025, ROM2F-2002-08, LPTENS-02-14, CPHT-RR-020-0202, CPHT-RR-020.0202",
    doi = "10.1016/S0370-1573(02)00273-9",
    journal = "Phys. Rept.",
    volume = "371",
    pages = "1--150",
    year = "2002",
    note = "[Erratum: Phys.Rept. 376, 407 (2003)]"
}

@inbook{nonsusystringsrevs2,
    author = "Angelantonj, Carlo and Florakis, Ioannis",
    title = "{A Lightning Introduction to String Theory}",
    eprint = "2406.09508",
    archivePrefix = "arXiv",
    primaryClass = "hep-th",
    doi = "10.1007/978-981-19-3079-9_53-1",
    year = "2024"
}

@article{nonsusystringsrevs3,
    author = "Leone, Giorgio and Raucci, Salvatore",
    title = "{Aspects of strings without spacetime supersymmetry}",
    eprint = "2509.24703",
    archivePrefix = "arXiv",
    primaryClass = "hep-th",
    reportNumber = "IFT-UAM/CSIC-25-100",
    month = "9",
    year = "2025"
}

@article{nonsusystringsrevs4,
    author = "Dudas, E. and Mourad, J. and Sagnotti, A.",
    title = "{Supersymmetry breaking with fields, strings and branes}",
    eprint = "2511.04367",
    archivePrefix = "arXiv",
    primaryClass = "hep-th",
    doi = "10.1016/j.physrep.2026.02.005",
    journal = "Phys. Rept.",
    volume = "1175",
    pages = "1--256",
    year = "2026"
}

@article{Dudas:2004nd,
    author = "Dudas, E. and Pradisi, G. and Nicolosi, M. and Sagnotti, A.",
    title = "{On tadpoles and vacuum redefinitions in string theory}",
    eprint = "hep-th/0410101",
    archivePrefix = "arXiv",
    reportNumber = "CPTH-RR-057-0904, LPT-ORSAY-04-82, ROM2F-04-28",
    doi = "10.1016/j.nuclphysb.2004.11.028",
    journal = "Nucl. Phys. B",
    volume = "708",
    pages = "3--44",
    year = "2005"
}

@article{Antoniadis:1998ki,
    author = "Antoniadis, Ignatios and Dudas, E. and Sagnotti, A.",
    title = "{Supersymmetry breaking, open strings and M theory}",
    eprint = "hep-th/9807011",
    archivePrefix = "arXiv",
    reportNumber = "CERN-TH-98-212, CPTH-S616-0698, LPTHE-ORSAY-98-42, ROM2F-98-21",
    doi = "10.1016/S0550-3213(98)00806-2",
    journal = "Nucl. Phys. B",
    volume = "544",
    pages = "469--502",
    year = "1999"
}

@article{Angelantonj:2007ts,
    author = "Angelantonj, Carlo and Dudas, Emilian",
    title = "{Metastable string vacua}",
    eprint = "0704.2553",
    archivePrefix = "arXiv",
    primaryClass = "hep-th",
    reportNumber = "CPHT-RR-017.0417, DFTT-2007-5, LPT-ORSAY-07-23",
    doi = "10.1016/j.physletb.2007.06.031",
    journal = "Phys. Lett. B",
    volume = "651",
    pages = "239--245",
    year = "2007"
}

@article{Mourad:2017rrl,
    author = "Mourad, J. and Sagnotti, A.",
    title = "{An Update on Brane Supersymmetry Breaking}",
    eprint = "1711.11494",
    archivePrefix = "arXiv",
    primaryClass = "hep-th",
    month = "11",
    year = "2017"
}

@article{Basile:2018irz,
    author = "Basile, I. and Mourad, J. and Sagnotti, A.",
    title = "{On Classical Stability with Broken Supersymmetry}",
    eprint = "1811.11448",
    archivePrefix = "arXiv",
    primaryClass = "hep-th",
    doi = "10.1007/JHEP01(2019)174",
    journal = "JHEP",
    volume = "01",
    pages = "174",
    year = "2019"
}

@article{Raucci:2022bjw,
    author = "Raucci, Salvatore",
    title = "{On new vacua of non-supersymmetric strings}",
    eprint = "2209.06537",
    archivePrefix = "arXiv",
    primaryClass = "hep-th",
    doi = "10.1016/j.physletb.2022.137663",
    journal = "Phys. Lett. B",
    volume = "837",
    pages = "137663",
    year = "2023"
}

@article{Basile:2021vxh,
    author = "Basile, Ivano",
    title = "{Supersymmetry breaking and stability in string vacua: Brane dynamics, bubbles and the swampland}",
    eprint = "2107.02814",
    archivePrefix = "arXiv",
    primaryClass = "hep-th",
    doi = "10.1007/s40766-021-00024-9",
    journal = "Riv. Nuovo Cim.",
    volume = "44",
    number = "10",
    pages = "499--596",
    year = "2021"
}

@article{Antonelli:2019nar,
    author = "Antonelli, Riccardo and Basile, Ivano",
    title = "{Brane annihilation in non-supersymmetric strings}",
    eprint = "1908.04352",
    archivePrefix = "arXiv",
    primaryClass = "hep-th",
    doi = "10.1007/JHEP11(2019)021",
    journal = "JHEP",
    volume = "11",
    pages = "021",
    year = "2019"
}

@article{Mourad:2021roa,
    author = "Mourad, J. and Sagnotti, A.",
    title = "{On warped string vacuum profiles and cosmologies. Part II. Non-supersymmetric strings}",
    eprint = "2109.12328",
    archivePrefix = "arXiv",
    primaryClass = "hep-th",
    doi = "10.1007/JHEP12(2021)138",
    journal = "JHEP",
    volume = "12",
    pages = "138",
    year = "2021"
}

@article{Mourad:2024mpg,
    author = "Mourad, J. and Raucci, S. and Sagnotti, A.",
    title = "{Brane profiles of non-supersymmetric strings}",
    eprint = "2406.16327",
    archivePrefix = "arXiv",
    primaryClass = "hep-th",
    doi = "10.1007/JHEP09(2024)019",
    journal = "JHEP",
    volume = "09",
    pages = "019",
    year = "2024"
}

@article{Angelantonj:2006ut,
    author = "Angelantonj, Carlo and Cardella, Matteo and Irges, Nikos",
    title = "{An Alternative for Moduli Stabilisation}",
    eprint = "hep-th/0608022",
    archivePrefix = "arXiv",
    reportNumber = "DFTT-16-2006, IFUM-869-FT",
    doi = "10.1016/j.physletb.2006.08.072",
    journal = "Phys. Lett. B",
    volume = "641",
    pages = "474--480",
    year = "2006"
}

\end{document}